\documentclass[aps,prl,reprint,amssymb,longbibliography,preprintnumbers,twocolumn,amsmath]{revtex4-2}
\usepackage{graphicx}
\usepackage{amsmath,bm}
\usepackage{hyperref}
\usepackage{physics}
\usepackage[final]{showlabels}

\begin{document}

\title{Propagation time and nondipole contributions to intraband high-harmonic generation}

\author{Simon Vendelbo Bylling Jensen}
\affiliation{Department of Physics and Astronomy, Aarhus
University, DK-8000 Aarhus C, Denmark}

\author{Lars Bojer Madsen}
\affiliation{Department of Physics and Astronomy, Aarhus
University, DK-8000 Aarhus C, Denmark}

\date{\today}

\begin{abstract}
Applying the semiclassical model, we identify two connected effects in intraband high-harmonic generation (HHG): (1) propagation time from the point of emission at the source to the point of detection, (2) beyond-electric-dipole corrections to the light-matter interaction. These effects inherit information regarding the dispersion and cause specific features in the spectra including even-ordered harmonics in systems with space- and time-inversion symmetry. They can for certain experimental geometries be measured free of the dipole background. 
\end{abstract}  
\maketitle

HHG is a central part of strong-field and attosecond science due to its capability to produce ultrafast bursts of coherent ultraviolet light and to probe ultrafast electron dynamics \cite{PhysRevA.66.023805,PhysRevLett.98.203007,doi:10.1126/science.1163077,doi:10.1126/science.1123904,PhysRevLett.94.053004,Itatani2004,Schubert2014,Garg2018,PhysRevLett.115.193603}. HHG in gases can be rationalized by the three-step model, in which an electron is ionized, then propagates through the external field, and later recombines upon emission of its excess energy \cite{PhysRevLett.68.3535,PhysRevA.49.2117,PhysRevLett.71.1994}. HHG within the diverse range of condensed matter systems is widely debated \cite{PhysRevA.85.043836,PhysRevA.91.013405,Lakhotia2020,PhysRevA.101.053411,PhysRevLett.124.153204,PhysRevLett.116.016601,PhysRevLett.113.073901,PhysRevLett.113.213901,PhysRevA.95.043416,PhysRevLett.122.193901,Hohenleutner2015,PhysRevA.94.063403,You2017}, but for bandgap materials often described by intra- and interband dynamics. The interband process bears similarities to the three-step model \cite{PhysRevLett.71.1994}, as the electron is excited to a conduction band, whereafter it propagates to later recombine with its hole and emit its excess energy \cite{PhysRevLett.113.073901}. For the intraband process, harmonics are generated as an electron wavepacket propagates through a band dispersion that deviates from the quadratic free-electron one. 
Although both processes are coupled \cite{PhysRevLett.113.073901,PhysRevB.77.075330}, generally the interband (intraband) mechanism dominates the HHG spectrum at harmonics beyond (below) the band gap. With increasing significance in the long-wavelength regime \cite{PhysRevB.91.064302}, the intraband process described with a semiclassical model has been excellent for modeling a variety of HHG experiments. It has reproduced characteristics of the spectrum and cutoff scaling of ZnO \cite{Ghimire2011}, orientation dependencies for MgO \cite{You2017}, orientation and polarization dependencies in GaSe \cite{PhysRevLett.120.243903}, qualitative features of the spectra of ZnS \cite{PhysRevB.102.104308}, polarization and bandstructure properties of ZnSe \cite{Lanin:17,Lanin:19}, qualitative dynamics of the Berry curvature for monolayer MoS$_2$ \cite{Liu2017}, spectral features and cutoff scaling for extreme ultraviolet beyond bandgap harmonics as well as reconstructed the Berry curvature of SiO$_2$ \cite{Luu2015,Luu2018}. In addition, interpretation of the semiclassical model is intuitive, as it bears resemblance to the equations of motion for free electrons driven by electric and magnetic fields. However, for free electrons in a similar strong-field long-wavelength regime, a breakdown of the electric dipole approximation for the light-matter interaction occurs in a variety of processes, see, e.g., Refs.~\cite{PhysRevLett.113.243001,Willenberg2019,PhysRevLett.123.093201,PhysRevLett.126.053202,PhysRevLett.106.193002,Hartung2019,Wang_2020,Haram_2020,Maurer_2021}. Beyond-dipole corrections are often associated with the high-frequency regime, and thus for condensed matter HHG, they have only been considered in the scope of examining the nondipole nature of the highest frequency components of the emitted harmonic field \cite{Gorlach2020}.  
An investigation of the validity of the dipole approximation within the low-frequency regime of the driving field seems crucial, as here the dipole approximation has been readily applied for the semiclassical approach in the literature \cite{Ghimire2011,PhysRevLett.120.243903,Liu2017,PhysRevB.102.104308,Lanin:19,Lanin:17,Luu2015,Luu2018}. Following this, identification of nondipole-induced features in the HHG spectra is critical in order to, e.g., distinguish them from topological features, which also arise in the semiclassical equations of motion \cite{Silva2019}. Furthermore, since a nondipole radiation-pressure force typically arises in the high-intensity long-wavelength regime  \cite{PhysRevA.48.R4027,PhysRevA.64.013411,PhysRevA.101.043408} it is relevant to reconsider the emitted radiation pattern since the emitter might be moving nontrivially towards or away from the detector. An unexplored characteristic of the intraband process is that the harmonics are emitted throughout the electron trajectory and not simply upon recombination. It differs significantly from the interpretation of gaseous or interband HHG, where emission occurs at the recombination step and the observed harmonics are affected by the phase-matching between multiple recombination sites (see the discussion in Supplemental Material \cite{SM}\nocite{Gaarde2008,PhysRevLett.81.297}). For intraband dynamics, however, there will be a time-dependent propagation time delay for the emitted harmonics to reach the detector. It depends on the instantaneous distance towards the detector for the wavepacket along its trajectory. This mechanism introduces new features in the HHG spectra that are connected to the nondipole-induced ones. Here we consider the following, (i) How does the variable propagation time delay from emission to observation of the harmonics affect the observed spectra? (ii) To what extent can nondipole effects alter the spectra? (iii) When are such effects important, and how are they related?

The semiclassical model for an electron wavepacket describes the dynamics of intraband electrons centered at position $\bm r$ with wavevector $\bm k$. We examine a system with space and time-inversion symmetry where the magnetization and Berry curvature vanish \cite{PhysRevB.59.14915}, allowing us to single out the behaviour stemming from propagation time and nondipole effects. The three-dimensional dispersion is expanded $\varepsilon(\textbf{k}) = \frac{\hbar^2}{4 a^2 m_e} \left\lbrace 1 + \sum_n \sum_{i=\lbrace x,y,z\rbrace} c_{n,i} \cos(n k_i a) \right\rbrace$ with material-dependent  $c_{n,i}$ coefficients. Similar $\varepsilon(\textbf{k})$'s are applied in the literature \cite{PhysRevA.85.043836,PhysRevB.91.064302,Ghimire2011,Liu2017,PhysRevB.102.104308,Lanin:19,Lanin:17,Luu2015,Luu2018}. Parameters of a Zinc-based crystal are applied with $c_{m,i} = -0.95 \delta_{m,1} -0.05 \delta_{m,3}$ and lattice spacing $a=5.4$ Å \cite{PhysRevB.102.104308}. The dynamics of the intraband wavepacket of charge $-e$, $e>0$, are governed by 
 \begin{equation}
\hbar \dot{\bm k} = -e\left( \bm E + \dot{\bm r} \times \bm{B}\right) \ \ \ \  \ \ \ \ \mathrm{and}  \ \ \ \ \ \ \ \  \dot{\bm r} = \frac{1}{\hbar}\pdv{\varepsilon(\bm k)}{\bm k}, \label{rdot}
\end{equation}
where $\bm E$  and $\bm B$ are the space- and time-dependent electric and magnetic fields derived from the vector potential $\bm{A} = A_0 ( \eta ) \left[0, \epsilon \cos(\eta), \sin(\eta) \right]^T$ with $ \eta = \omega t - \omega x/c$ where $c$ is the speed of light. Throughout this work an angular frequency of $\omega = 0.0217$  a.u. ($\lambda = 2100$ nm) is applied. $A_0 ( \eta )$ is a $30$-cycle $\sin^2$ envelope function, and $\epsilon \in [0;1]$ is describing the polarization, with $\epsilon = 0$ and $\epsilon = 1$ for linearly and circularly polarized light, respectively. To retain an intensity of $I = 5.3 \times 10^{11}$ W/cm$^2$ when varying $\epsilon$, the amplitude of the vector potential in atomic units is scaled as $A_0 = \sqrt{I}/\omega \sqrt{1+\epsilon^2}$. When considering nondipole effects within the strong-field, long-wavelength limit, it is useful to expand the vector potential as $\bm A = \sum_{l=0}^\infty \bm A^{(l)}$ with $\bm A^{(l)}$ denoting the $l$'th order of $\omega x/c$ such that $\bm A^{(l)} = (l!)^{-1}  \dv*[l]{\bm A}{\eta} \eval{}_{\eta = \omega t} \left(-\omega x/c\right)^l$. The electric and magnetic fields inherit the notation of the associated $\bm A^{(l)}$, such that $\bm E^{(l)}=-\partial_t \bm A^{(l)}$ and $\bm B^{(l)}=\nabla \times \bm A^{(l)}$. Thereby $\bm A^{(0)}$ is the term within the commonly applied electric dipole approximation, in which the term $ q \dot{\bm r} \times \bm{B}$ of Eq. \eqref{rdot} vanishes as $\bm B^{(0)}=0$. 

\begin{figure}
\includegraphics[width=8.6 cm]{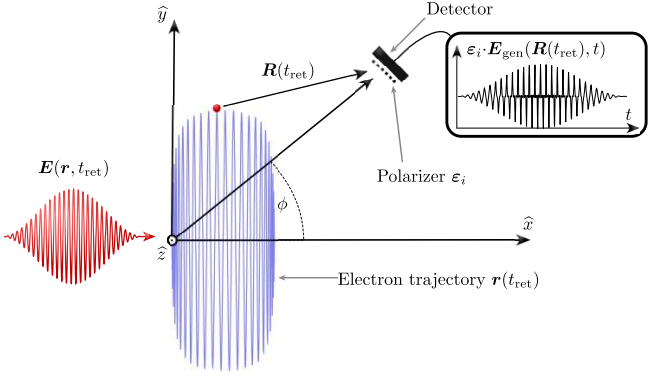}
\caption{Illustration of the electron dynamics in retarded time. The incoming electric field $\bm E(\bm r,t_{\text{ret}})$, polarized in the $(y,z)$-plane and propagating in the $x$-direction, induces an electron wavepacket trajectory, which is magnified in the figure. Relative to this trajectory, the detector position is noted $\bm R(t_{\text{ret}})$ and depends on the retarded emission time $t_{\text{ret}}$. After the propagational delay between $t_{\text{ret}}$ and the detection time $t$, the detector, placed at angle $\phi$ in the $(x,y)$-plane, measures the spectrum of the generated field $\bm E_{\text{gen}}(\bm R (t_{\text{ret}}),t)$ with an associated polarizer singling out $\bm \varepsilon_i$-polarized harmonics.} \label{fig:1}
\end{figure}
For thin targets, where phasematching effects can be neglected \cite{PhysRevLett.125.083901}, the acceleration electric field emitted by the electronic wavepacket, observed at $\bm R (t_{\text{ret}})$ is \cite{jackson_classical_1999} 
\begin{equation}
\bm{E}_{\text{gen}}(\bm R  (t_{\text{ret}}), t) =  -\frac{ e \bm R  (t_{\text{ret}}) \times \left[  \bm R  (t_{\text{ret}}) \times  \ddot{\bm r} (t_{\text{ret}}) \right]}{ 4\pi\epsilon_0 c^2 R (t_{\text{ret}})^3} .
 \label{eq:new}
\end{equation}
For an illustration of the geometry see Fig.~\ref{fig:1}. The harmonics generated at the retarded emission time $t_{\text{ret}}$ are observed at the detection time $t=t_{\text{ret}} +  R(t_{\text{ret}})/c$ after the propagation time delay $R(t_{\text{ret}})/c $. The electron trajectory is therefore evaluated at the retarded time. Typically for HHG in solids, an approximation to this formula is applied, in which one neglects propagational time delay, i.e., the difference between the time of emission $t_{\text{ret}}$ and the time of detection $t$. In doing so, the emitted spectra are derived from the current $\bm j(t)$ or the time derivative hereof, which for the case of a localized electronic wavepacket, can be written as
\begin{equation}
\bm{E}_{\text{gen},0}(t) \propto -\dv{\bm{j}(t)}{t} = e\ddot{\bm{r}}(t).  \label{eq:old}
\end{equation}
In general, we consider a detector placed at an angle $\phi$ in the $(x,y)$-plane measuring $\bm \varepsilon_i$-polarized light at a distance of $1$ m from the origin of the electronic wavepacket [Fig.~\ref{fig:1}]. We have checked that the results of the present work are independent of the macroscopic distance $R(t = 0)$. For the numerical simulation, we initialize an electron wavepacket at the $\Gamma$-point, $\bm k (t=0) = [0,0,0]^T$ and $\bm r (t=0) = [0,0,0]^T$, similar to Refs.\cite{PhysRevB.102.104308,Lanin:17,Luu2015,Luu2018}. The model does not account for field-induced electron correlation effects which is justified when applying a single driving pulse \cite{PhysRevB.104.054309}. The electron wavepacket is propagated through Eq.~\eqref{rdot}, beyond and within the dipole approximation. 
\begin{figure}
\includegraphics[width=8.6 cm]{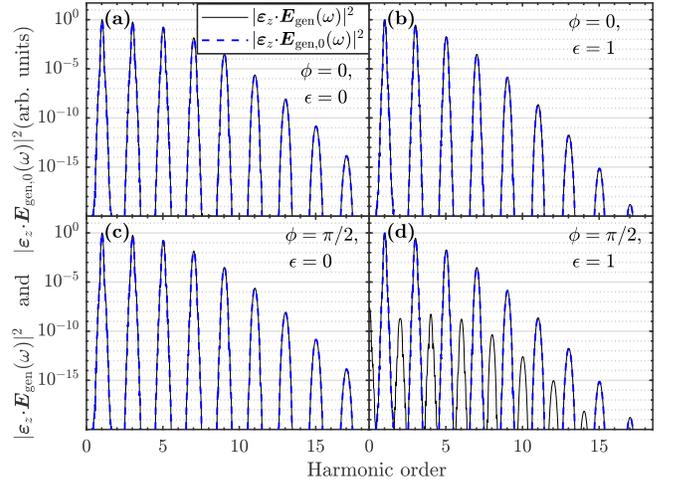}
\caption{Calculated $\bm{\varepsilon}_z$-polarized HHG spectra observed at $\phi = 0$ (see Fig.~\ref{fig:1}) with (a) linearly polarized and (b) circularly polarized driving fields. Similar spectra are given with $\phi = \pi/2$ for (c) and (d), respectively. The radiation pattern  $|E_{\text{gen}}(\omega)|^2$ from  Eq.~\eqref{eq:new} (black continuous) is compared to the approximation $|E_{\text{gen},0}(\omega)|^2$ of Eq.~\eqref{eq:old} (blue dashed) induced by an intraband electronic wavepacket with initial conditions $\bm r = [0,0,0]^T$ and $\bm k = [0,0,0]^T$. See text for laser parameters.} \label{fig:2}
\end{figure}

First, in Fig~\ref{fig:2}, we investigate solely the propagation time delay effect. To this end, we consider a regime where nondipole effects are shown later to be negligible, namely by considering $\bm \varepsilon_z$-polarized harmonics originating dominantly from the $z$-polarized current within the polarization plane. In Fig.~\ref{fig:2}, we compare the predictions of Eq.~\eqref{eq:new} with those of the approximate Eq.~\eqref{eq:old}, for a linearly and circularly polarized field with detector positions $\phi = \lbrace 0, \pi/2 \rbrace$. Figure~\ref{fig:2}(a) shows that the approximation of Eq.~\eqref{eq:old} is justified at $\phi = 0$. The reason being that the temporal variation in $\bm R(t_{\text{ret}})$ is relatively small when the detector is placed perpendicular to the polarization direction of the driving field. In Fig.~\ref{fig:2}(b) the electron dynamics is no longer one-dimensional, as the wavepacket is driven in the polarization $(y,z)$-plane of the circularly polarized driving field. Since, however, the detector is placed in a direction perpendicular to this plane, Eq.~\eqref{eq:old} is still accurate. Comparing Figs.~\ref{fig:2}(a) and (b), we note that a reduction is found in the spectra with increasing ellipticity, due to the amplitude of the electric vector potential decreasing for fixed intensity. The electron driven linearly in the $z$-direction will emit a symmetrical radiation pattern in the $(x,y)$-plane, as seen when comparing different $\phi$ in Figs.~\ref{fig:2}(a) and (c). For the circularly polarized driving field, however, this symmetry is broken by propagation time delay as observed in Fig.~\ref{fig:2}(d). For finite $\phi$ the electron has an excursion in the direction of the detector, causing a relatively larger change in the distance to the detector $R(t_{\text{ret}})$, and a time-dependent propagation delay [Fig.~\ref{fig:1}]. This effect infers that an electron accelerating with a given frequency $l\omega$ will be observed to attain additional frequency components of $(l \pm 1) \omega$. 
In Fig.~\ref{fig:2}(d) this is apparent as even harmonics arise with intensity scaling as $\sin^2(\phi)$. Propagation delay effects can seemingly be neglected in Figs.~\ref{fig:2}(a), (b), and (c), where the electron trajectory is confined to a plane perpendicular to the direction towards the detector and the effect of the variations of the propagational delay is of higher order. Oppositely such effects must be included if the detector occupies a solid angle with multiple $\phi$ components, or if a significant excursion of the wavepacket is in the direction of the detector, where it generates additional side peaks in the spectra. The magnitude of such side peaks depends on the electron trajectory, and can thus bring insight into properties of the material.

\begin{figure}
\includegraphics[width=8.6 cm]{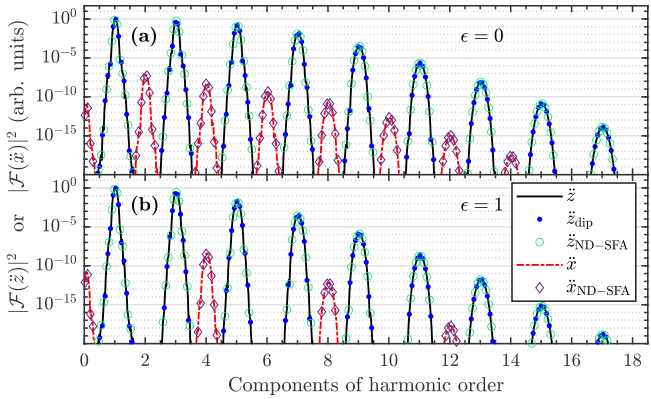}
\caption{Norm square of the Fourier transform of the acceleration along the polarization $\ddot{z}$ or propagation $\ddot{x}$ direction for a driving field with (a) linear $\epsilon = 0$ and (b) circular $\epsilon = 1$ polarization. The acceleration induced by the electromagnetic field, is compared to that induced by the dipole field and the $\text{ND-SFA}$ approach of Eq.~\eqref{ndsfa}. For the propagation direction, the acceleration obtained within the dipole approximation vanishes. The parameters are as in Fig.~\ref{fig:2}.} \label{fig:3}
\end{figure}

We now analyse nondipole effects in Fig.~\ref{fig:3}. In order to single these out from the propagation delay effects, we start by analyzing the acceleration of the electron wavepacket to characterize the nondipole components hereof. Later the nondipole effects on the HHG spectrum will be examined inserting the acceleration in Eq.~\eqref{eq:new}. In regions where Eq.~\eqref{eq:old} is accurate, we can draw conclusions based on the frequency components of the wavepacket acceleration. To investigate the nondipole corrections to the electron acceleration we include leading-order effects from the interaction between the dipole-induced motion and the magnetic field. A similar approach was found to account for nondipole effects in atomic systems \cite{PhysRevA.101.043408,Jensen2020,Lund2021}. In this nondipole strong-field approximation (ND-SFA) approach, the first part of Eq.~\eqref{rdot} reduces to 
 \begin{equation}
\hbar \dot{\bm k} = -e \left[ \left(- \partial_t \bm A^{(0)} \right) + \left(\dot{r}_y B_z^{(1)}-\dot{r}_z B_y^{(1)} \right)\widehat{x}\right], \label{ndsfa}
\end{equation}
where the last term describes a radiation-pressure-like force on the wavepacket in the laser propagation direction \cite{PhysRevA.101.043408}. Simulations with the fully retarded, the ND-SFA and the dipole field are compared in Fig.~\ref{fig:3}. For the acceleration within the polarization plane [$\ddot{z}$ in Figs.~\ref{fig:3}(a) and (b)] the dynamics are well described within the dipole approximation, consistent with the free-electron case \cite{PhysRevA.101.043408}. Also, similar to the free-electron case, significant nondipole corrections arise for the dynamics in the propagation direction in Figs.~\ref{fig:3}(a) and (b) where a harmonic nondipole acceleration containing even-ordered multiples of the driving frequency is observed. Remarkably for circularly-polarized light, in Fig.~\ref{fig:3}(b), only fourth-order multiples are observed. In the long-pulse circular-polarized-driving-field regime, a free electron would have $\dot{r}_y \propto A^{(0)}_y$ and $\dot{r}_z \propto A^{(0)}_z$ in which case the last term of Eq.~\eqref{ndsfa} vanishes, as $\bm A^{(0)} \parallel \bm B^{(1)}$ \cite{PhysRevA.101.043408}. This is not the case for the intraband wavepacket as its induced velocity in the polarization plane is not proportional to $\bm A^{(0)}$, but modified by the non-parabolic band dispersion as captured by the last part of Eq.~\eqref{rdot}. These predictions solidify the different nature of nondipole effects for atomic or intraband dynamics. We observe in Fig.~\ref{fig:3} that the ND-SFA approach accurately describes these dynamics. Due to its simplicity, it is suitable for analytical calculations (see Supplemental Material \cite{SM}) in which one can find the wavepacket acceleration to be
\begin{widetext}
\begin{equation}
\ddot{\bm r} = \frac{\hbar \omega}{2 a m_e}  \sum_n  \sum_{l=1,3,5,...}^\infty \begin{bmatrix}  -\frac{\hbar (l+1)}{a m^\star c}  \left\lbrace   c_{n,z}  J_{l+1} \left(\frac{n a e}{\hbar}A_0(\omega t) \right) +  c_{n,y} (-1)^{(l+1)/2} J_{l+1} \left(\frac{n a e}{\hbar}\epsilon A_0(\omega t) \right) \right\rbrace  \sin((l+1) \omega t) \\ nl  c_{n,y} (-1)^{(l-1)/2}  J_{l} (\frac{n a e }{\hbar} \epsilon A_0(\omega t)) \sin(l \omega t) \\ -nl  c_{n,z} J_{l} (\frac{n a e}{\hbar} A_0(\omega t)) \cos(l\omega t) \end{bmatrix}, \label{eq:analytical}
\end{equation}
\end{widetext}
where $J_l$ are Bessel functions and $m^{\star}$ is the effective mass of the dispersion along the propagation direction \cite{SM}. We note that the arguments of the Bessel functions in the long pulse limit can be expressed in terms of $\omega_B / \omega$ with $\omega_B = eaA_0\omega/\hbar$ the Bloch frequency. The analytical approach accurately describes the dominant nondipole dynamics, and allows for characterizing the selection rules for the harmonic contributions to the acceleration \cite{SM}. Generally, with space- and time-inversion symmetry only odd (even) frequency components will be present in the $y$- and $z$-direction ($x$-direction), respectively \cite{SM}. Furthermore for circularly polarized driving field and if the dispersion in both directions of the polarization plane is identical, $c_{n,y}=c_{n,z}$, then the $x$-direction will consist of only fourth-ordered harmonic motion  \cite{SM}, similar to what is observed in Fig.~\ref{fig:3}(b). The detection of the nondipole acceleration thus provides a tool to investigate the symmetries of the material at hand as well as the dispersion along both the propagation and polarization directions.     

\begin{figure}
\includegraphics[width=8.6 cm]{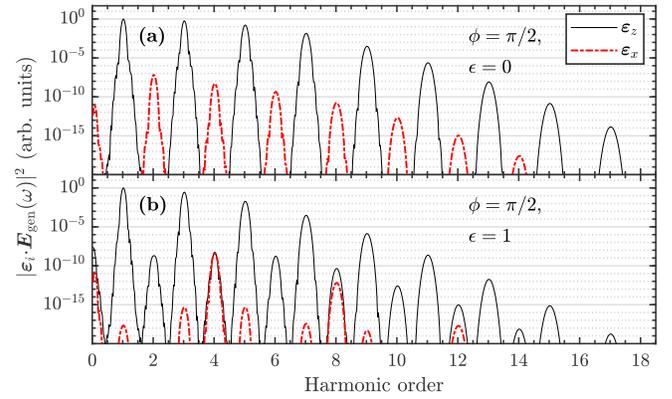}
\caption{HHG spectra, $| \bm \varepsilon_i \cdot \bm E_{\text{gen}}(\omega)|^2$, including polarization time delay in Eq.~\eqref{eq:new} for the $\ddot{z}$- and $\ddot{x}$-accelerations in Fig.~\ref{fig:3}. The detector is placed at an angle of $\phi = \pi/2$.} \label{fig:4}
\end{figure}

Lastly, we consider in Fig.~\ref{fig:4} the interplay between the propagation time and nondipole effects. In general the two mechanisms couple since nondipole corrections to the electron trajectory modify $\bm R (t_{\text{ret}})$ of Eq.~\eqref{eq:new} altering the propogation time delay $R(t_{\text{ret}})/c$. Furthermore, the nondipole $\bm{\varepsilon}_z$-polarized harmonics are sensitive to propagation time delay as they are observed at nonvanishing $\phi$. We consider a detector placed at $\phi = \pi/2$ since here both the emitted HHG spectra of the electron acceleration from the polarization plane and from the nondipole $x$-motion can be observed by considering the $\bm{\varepsilon}_z$- and $\bm{\varepsilon}_x$-polarized harmonics, respectively. The HHG spectra, calculated from Eq.~\eqref{eq:new}, are given in Fig.~\ref{fig:4}. At first, when comparing Fig.~\ref{fig:3}(a) and Fig.~\ref{fig:4}(a), no significant propagation delay effects are observed. This is due to the nondipole-induced dynamics along the propagation direction being of order $1/c$ lower than the excursion in the polarization plane. Similarly, no nondipole effects were found for the spectra of Fig.~\ref{fig:2}(a) and (c). For circularly polarized light, the excursion in the $y$-direction of the polarization plane contributes with additional side peaks to Fig.~\ref{fig:4}(b), when compared to Fig.~\ref{fig:3}(b), for both the $\bm \varepsilon_z$-polarized and the nondipole $\bm \varepsilon_x$-polarized harmonics. Interestingly for the fourth harmonic in Fig.~\ref{fig:4}(b), the propagation time delay side peak from the $\bm \varepsilon_z$-polarized harmonics, is of similar magnitude as the dominant nondipole harmonic with $\bm \varepsilon_x$-polarization. The interplay of nondipole effects and propagation delay effects can thus be investigated by considering the ellipticity-dependence of specific harmonics as a function of $\phi$ and the signatures of each mechanism can be readily distinguished in experiments by varying the detector angle $\phi$ and observed polarization $\bm \varepsilon_i$. If observing $\bm \varepsilon_x$-polarized harmonics at $\phi = \pi/2$, nondipole harmonics can be identified free of the vanishing background of harmonics emitted within the dipole approximation. Both propagation time and nondipole effects can thus be singled out by varying the driving field ellipticity and detection angle. Features arising in the HHG spectra from both mechanisms carry information about the dispersion both within the polarization and propagation directions. Similarly, the nondipole harmonics are sensitive to the symmetry of the sample, as discussed in relation to Fig.~\ref{fig:3}, and therefore in principle provide a path to examining structural changes on an ultrafast timescale. We also note that the significance of nondipole harmonics, as well as propagation time delay contributions, is restricted by the effective mass of the system [see Eq.~\eqref{eq:analytical}], and is expected to increase with lower effective mass, where also the harmonic cutoff is expected to be extended \cite{PhysRevB.84.081202,PhysRevB.77.075330,Golde2011}.

In conclusion, by including propagation time delay to intraband HHG, additional side peaks appear in the HHG spectra, resulting in even harmonics despite space- and time-inversion symmetries in the sample. The propagation delay depends on the position of the detector compared to the emitting electron wavepacket and the effect is negligible if the wavepacket is having a relatively small amplitude motion in the direction towards the detector. We have characterized nondipole corrections to the electron wavepacket dynamics as an oscillatory motion in the propagation direction of the driving field. By analyzing a leading-order approximation to the equation of motion, we provide an analytical assessment of the nondipole effects [Eq.~\eqref{eq:analytical}]. For inversion symmetric lattices, selection rules are found for the emitted nondipole harmonics. These are even- or fourth-ordered for the case of a linearly or circularly polarized driving field. For a complete description of the HHG process, one must be aware of the interplay of propagation time and nondipole effects. Both effects bring information regarding material properties, which affect the real space electronic trajectory and velocity. Similarly, they provide a tool for investigating the symmetries of the sample and reconstructing the dispersion both within and out of the polarization plane.

\begin{acknowledgments}
We thank Stefano M. Cavaletto for useful suggestions. This work was supported by the Danish Council for Independent Research (GrantNo.9040-00001B).
\end{acknowledgments}


\begin{thebibliography}{60}%
\makeatletter
\providecommand \@ifxundefined [1]{%
 \@ifx{#1\undefined}
}%
\providecommand \@ifnum [1]{%
 \ifnum #1\expandafter \@firstoftwo
 \else \expandafter \@secondoftwo
 \fi
}%
\providecommand \@ifx [1]{%
 \ifx #1\expandafter \@firstoftwo
 \else \expandafter \@secondoftwo
 \fi
}%
\providecommand \natexlab [1]{#1}%
\providecommand \enquote  [1]{``#1''}%
\providecommand \bibnamefont  [1]{#1}%
\providecommand \bibfnamefont [1]{#1}%
\providecommand \citenamefont [1]{#1}%
\providecommand \href@noop [0]{\@secondoftwo}%
\providecommand \href [0]{\begingroup \@sanitize@url \@href}%
\providecommand \@href[1]{\@@startlink{#1}\@@href}%
\providecommand \@@href[1]{\endgroup#1\@@endlink}%
\providecommand \@sanitize@url [0]{\catcode `\\12\catcode `\$12\catcode
  `\&12\catcode `\#12\catcode `\^12\catcode `\_12\catcode `\%12\relax}%
\providecommand \@@startlink[1]{}%
\providecommand \@@endlink[0]{}%
\providecommand \url  [0]{\begingroup\@sanitize@url \@url }%
\providecommand \@url [1]{\endgroup\@href {#1}{\urlprefix }}%
\providecommand \urlprefix  [0]{URL }%
\providecommand \Eprint [0]{\href }%
\providecommand \doibase [0]{https://doi.org/}%
\providecommand \selectlanguage [0]{\@gobble}%
\providecommand \bibinfo  [0]{\@secondoftwo}%
\providecommand \bibfield  [0]{\@secondoftwo}%
\providecommand \translation [1]{[#1]}%
\providecommand \BibitemOpen [0]{}%
\providecommand \bibitemStop [0]{}%
\providecommand \bibitemNoStop [0]{.\EOS\space}%
\providecommand \EOS [0]{\spacefactor3000\relax}%
\providecommand \BibitemShut  [1]{\csname bibitem#1\endcsname}%
\let\auto@bib@innerbib\@empty
\bibitem [{\citenamefont {Lein}\ \emph {et~al.}(2002)\citenamefont {Lein},
  \citenamefont {Hay}, \citenamefont {Velotta}, \citenamefont {Marangos},\ and\
  \citenamefont {Knight}}]{PhysRevA.66.023805}%
  \BibitemOpen
  \bibfield  {author} {\bibinfo {author} {\bibfnamefont {M.}~\bibnamefont
  {Lein}}, \bibinfo {author} {\bibfnamefont {N.}~\bibnamefont {Hay}}, \bibinfo
  {author} {\bibfnamefont {R.}~\bibnamefont {Velotta}}, \bibinfo {author}
  {\bibfnamefont {J.~P.}\ \bibnamefont {Marangos}},\ and\ \bibinfo {author}
  {\bibfnamefont {P.~L.}\ \bibnamefont {Knight}},\ }\bibfield  {title}
  {\bibinfo {title} {Interference effects in high-order harmonic generation
  with molecules},\ }\href {https://doi.org/10.1103/PhysRevA.66.023805}
  {\bibfield  {journal} {\bibinfo  {journal} {Phys. Rev. A}\ }\textbf {\bibinfo
  {volume} {66}},\ \bibinfo {pages} {023805} (\bibinfo {year}
  {2002})}\BibitemShut {NoStop}%
\bibitem [{\citenamefont {Torres}\ \emph {et~al.}(2007)\citenamefont {Torres},
  \citenamefont {Kajumba}, \citenamefont {Underwood}, \citenamefont {Robinson},
  \citenamefont {Baker}, \citenamefont {Tisch}, \citenamefont {de~Nalda},
  \citenamefont {Bryan}, \citenamefont {Velotta}, \citenamefont {Altucci},
  \citenamefont {Turcu},\ and\ \citenamefont
  {Marangos}}]{PhysRevLett.98.203007}%
  \BibitemOpen
  \bibfield  {author} {\bibinfo {author} {\bibfnamefont {R.}~\bibnamefont
  {Torres}}, \bibinfo {author} {\bibfnamefont {N.}~\bibnamefont {Kajumba}},
  \bibinfo {author} {\bibfnamefont {J.~G.}\ \bibnamefont {Underwood}}, \bibinfo
  {author} {\bibfnamefont {J.~S.}\ \bibnamefont {Robinson}}, \bibinfo {author}
  {\bibfnamefont {S.}~\bibnamefont {Baker}}, \bibinfo {author} {\bibfnamefont
  {J.~W.~G.}\ \bibnamefont {Tisch}}, \bibinfo {author} {\bibfnamefont
  {R.}~\bibnamefont {de~Nalda}}, \bibinfo {author} {\bibfnamefont {W.~A.}\
  \bibnamefont {Bryan}}, \bibinfo {author} {\bibfnamefont {R.}~\bibnamefont
  {Velotta}}, \bibinfo {author} {\bibfnamefont {C.}~\bibnamefont {Altucci}},
  \bibinfo {author} {\bibfnamefont {I.~C.~E.}\ \bibnamefont {Turcu}},\ and\
  \bibinfo {author} {\bibfnamefont {J.~P.}\ \bibnamefont {Marangos}},\
  }\bibfield  {title} {\bibinfo {title} {Probing orbital structure of
  polyatomic molecules by high-order harmonic generation},\ }\href
  {https://doi.org/10.1103/PhysRevLett.98.203007} {\bibfield  {journal}
  {\bibinfo  {journal} {Phys. Rev. Lett.}\ }\textbf {\bibinfo {volume} {98}},\
  \bibinfo {pages} {203007} (\bibinfo {year} {2007})}\BibitemShut {NoStop}%
\bibitem [{\citenamefont {Li}\ \emph {et~al.}(2008)\citenamefont {Li},
  \citenamefont {Zhou}, \citenamefont {Lock}, \citenamefont {Patchkovskii},
  \citenamefont {Stolow}, \citenamefont {Kapteyn},\ and\ \citenamefont
  {Murnane}}]{doi:10.1126/science.1163077}%
  \BibitemOpen
  \bibfield  {author} {\bibinfo {author} {\bibfnamefont {W.}~\bibnamefont
  {Li}}, \bibinfo {author} {\bibfnamefont {X.}~\bibnamefont {Zhou}}, \bibinfo
  {author} {\bibfnamefont {R.}~\bibnamefont {Lock}}, \bibinfo {author}
  {\bibfnamefont {S.}~\bibnamefont {Patchkovskii}}, \bibinfo {author}
  {\bibfnamefont {A.}~\bibnamefont {Stolow}}, \bibinfo {author} {\bibfnamefont
  {H.~C.}\ \bibnamefont {Kapteyn}},\ and\ \bibinfo {author} {\bibfnamefont
  {M.~M.}\ \bibnamefont {Murnane}},\ }\bibfield  {title} {\bibinfo {title}
  {Time-resolved dynamics in {{N$_2$O$_4$}} probed using high harmonic
  generation},\ }\href {https://doi.org/10.1126/science.1163077} {\bibfield
  {journal} {\bibinfo  {journal} {Science}\ }\textbf {\bibinfo {volume}
  {322}},\ \bibinfo {pages} {1207} (\bibinfo {year} {2008})}\BibitemShut
  {NoStop}%
\bibitem [{\citenamefont {Baker}\ \emph {et~al.}(2006)\citenamefont {Baker},
  \citenamefont {Robinson}, \citenamefont {Haworth}, \citenamefont {Teng},
  \citenamefont {Smith}, \citenamefont {Chirilă}, \citenamefont {Lein},
  \citenamefont {Tisch},\ and\ \citenamefont
  {Marangos}}]{doi:10.1126/science.1123904}%
  \BibitemOpen
  \bibfield  {author} {\bibinfo {author} {\bibfnamefont {S.}~\bibnamefont
  {Baker}}, \bibinfo {author} {\bibfnamefont {J.~S.}\ \bibnamefont {Robinson}},
  \bibinfo {author} {\bibfnamefont {C.~A.}\ \bibnamefont {Haworth}}, \bibinfo
  {author} {\bibfnamefont {H.}~\bibnamefont {Teng}}, \bibinfo {author}
  {\bibfnamefont {R.~A.}\ \bibnamefont {Smith}}, \bibinfo {author}
  {\bibfnamefont {C.~C.}\ \bibnamefont {Chirilă}}, \bibinfo {author}
  {\bibfnamefont {M.}~\bibnamefont {Lein}}, \bibinfo {author} {\bibfnamefont
  {J.~W.~G.}\ \bibnamefont {Tisch}},\ and\ \bibinfo {author} {\bibfnamefont
  {J.~P.}\ \bibnamefont {Marangos}},\ }\bibfield  {title} {\bibinfo {title}
  {Probing proton dynamics in molecules on an attosecond time scale},\ }\href
  {https://doi.org/10.1126/science.1123904} {\bibfield  {journal} {\bibinfo
  {journal} {Science}\ }\textbf {\bibinfo {volume} {312}},\ \bibinfo {pages}
  {424} (\bibinfo {year} {2006})}\BibitemShut {NoStop}%
\bibitem [{\citenamefont {Lein}(2005)}]{PhysRevLett.94.053004}%
  \BibitemOpen
  \bibfield  {author} {\bibinfo {author} {\bibfnamefont {M.}~\bibnamefont
  {Lein}},\ }\bibfield  {title} {\bibinfo {title} {Attosecond probing of
  vibrational dynamics with high-harmonic generation},\ }\href
  {https://doi.org/10.1103/PhysRevLett.94.053004} {\bibfield  {journal}
  {\bibinfo  {journal} {Phys. Rev. Lett.}\ }\textbf {\bibinfo {volume} {94}},\
  \bibinfo {pages} {053004} (\bibinfo {year} {2005})}\BibitemShut {NoStop}%
\bibitem [{\citenamefont {Itatani}\ \emph {et~al.}(2004)\citenamefont
  {Itatani}, \citenamefont {Levesque}, \citenamefont {Zeidler}, \citenamefont
  {Niikura}, \citenamefont {P{\'e}pin}, \citenamefont {Kieffer}, \citenamefont
  {Corkum},\ and\ \citenamefont {Villeneuve}}]{Itatani2004}%
  \BibitemOpen
  \bibfield  {author} {\bibinfo {author} {\bibfnamefont {J.}~\bibnamefont
  {Itatani}}, \bibinfo {author} {\bibfnamefont {J.}~\bibnamefont {Levesque}},
  \bibinfo {author} {\bibfnamefont {D.}~\bibnamefont {Zeidler}}, \bibinfo
  {author} {\bibfnamefont {H.}~\bibnamefont {Niikura}}, \bibinfo {author}
  {\bibfnamefont {H.}~\bibnamefont {P{\'e}pin}}, \bibinfo {author}
  {\bibfnamefont {J.~C.}\ \bibnamefont {Kieffer}}, \bibinfo {author}
  {\bibfnamefont {P.~B.}\ \bibnamefont {Corkum}},\ and\ \bibinfo {author}
  {\bibfnamefont {D.~M.}\ \bibnamefont {Villeneuve}},\ }\bibfield  {title}
  {\bibinfo {title} {Tomographic imaging of molecular orbitals},\ }\href
  {https://doi.org/10.1038/nature03183} {\bibfield  {journal} {\bibinfo
  {journal} {Nature}\ }\textbf {\bibinfo {volume} {432}},\ \bibinfo {pages}
  {867} (\bibinfo {year} {2004})}\BibitemShut {NoStop}%
\bibitem [{\citenamefont {Schubert}\ \emph {et~al.}(2014)\citenamefont
  {Schubert}, \citenamefont {Hohenleutner}, \citenamefont {Langer},
  \citenamefont {Urbanek}, \citenamefont {Lange}, \citenamefont {Huttner},
  \citenamefont {Golde}, \citenamefont {Meier}, \citenamefont {Kira},
  \citenamefont {Koch},\ and\ \citenamefont {Huber}}]{Schubert2014}%
  \BibitemOpen
  \bibfield  {author} {\bibinfo {author} {\bibfnamefont {O.}~\bibnamefont
  {Schubert}}, \bibinfo {author} {\bibfnamefont {M.}~\bibnamefont
  {Hohenleutner}}, \bibinfo {author} {\bibfnamefont {F.}~\bibnamefont
  {Langer}}, \bibinfo {author} {\bibfnamefont {B.}~\bibnamefont {Urbanek}},
  \bibinfo {author} {\bibfnamefont {C.}~\bibnamefont {Lange}}, \bibinfo
  {author} {\bibfnamefont {U.}~\bibnamefont {Huttner}}, \bibinfo {author}
  {\bibfnamefont {D.}~\bibnamefont {Golde}}, \bibinfo {author} {\bibfnamefont
  {T.}~\bibnamefont {Meier}}, \bibinfo {author} {\bibfnamefont
  {M.}~\bibnamefont {Kira}}, \bibinfo {author} {\bibfnamefont {S.~W.}\
  \bibnamefont {Koch}},\ and\ \bibinfo {author} {\bibfnamefont
  {R.}~\bibnamefont {Huber}},\ }\bibfield  {title} {\bibinfo {title} {Sub-cycle
  control of terahertz high-harmonic generation by dynamical {{B}}loch
  oscillations},\ }\href {https://doi.org/10.1038/nphoton.2013.349} {\bibfield
  {journal} {\bibinfo  {journal} {Nature Photonics}\ }\textbf {\bibinfo
  {volume} {8}},\ \bibinfo {pages} {119} (\bibinfo {year} {2014})}\BibitemShut
  {NoStop}%
\bibitem [{\citenamefont {Garg}\ \emph {et~al.}(2018)\citenamefont {Garg},
  \citenamefont {Kim},\ and\ \citenamefont {Goulielmakis}}]{Garg2018}%
  \BibitemOpen
  \bibfield  {author} {\bibinfo {author} {\bibfnamefont {M.}~\bibnamefont
  {Garg}}, \bibinfo {author} {\bibfnamefont {H.~Y.}\ \bibnamefont {Kim}},\ and\
  \bibinfo {author} {\bibfnamefont {E.}~\bibnamefont {Goulielmakis}},\
  }\bibfield  {title} {\bibinfo {title} {Ultimate waveform reproducibility of
  extreme-ultraviolet pulses by high-harmonic generation in quartz},\ }\href
  {https://doi.org/10.1038/s41566-018-0123-6} {\bibfield  {journal} {\bibinfo
  {journal} {Nature Photonics}\ }\textbf {\bibinfo {volume} {12}},\ \bibinfo
  {pages} {291} (\bibinfo {year} {2018})}\BibitemShut {NoStop}%
\bibitem [{\citenamefont {Vampa}\ \emph
  {et~al.}(2015{\natexlab{a}})\citenamefont {Vampa}, \citenamefont {Hammond},
  \citenamefont {Thir\'e}, \citenamefont {Schmidt}, \citenamefont {L\'egar\'e},
  \citenamefont {McDonald}, \citenamefont {Brabec}, \citenamefont {Klug},\ and\
  \citenamefont {Corkum}}]{PhysRevLett.115.193603}%
  \BibitemOpen
  \bibfield  {author} {\bibinfo {author} {\bibfnamefont {G.}~\bibnamefont
  {Vampa}}, \bibinfo {author} {\bibfnamefont {T.~J.}\ \bibnamefont {Hammond}},
  \bibinfo {author} {\bibfnamefont {N.}~\bibnamefont {Thir\'e}}, \bibinfo
  {author} {\bibfnamefont {B.~E.}\ \bibnamefont {Schmidt}}, \bibinfo {author}
  {\bibfnamefont {F.}~\bibnamefont {L\'egar\'e}}, \bibinfo {author}
  {\bibfnamefont {C.~R.}\ \bibnamefont {McDonald}}, \bibinfo {author}
  {\bibfnamefont {T.}~\bibnamefont {Brabec}}, \bibinfo {author} {\bibfnamefont
  {D.~D.}\ \bibnamefont {Klug}},\ and\ \bibinfo {author} {\bibfnamefont
  {P.~B.}\ \bibnamefont {Corkum}},\ }\bibfield  {title} {\bibinfo {title}
  {All-optical reconstruction of crystal band structure},\ }\href
  {https://doi.org/10.1103/PhysRevLett.115.193603} {\bibfield  {journal}
  {\bibinfo  {journal} {Phys. Rev. Lett.}\ }\textbf {\bibinfo {volume} {115}},\
  \bibinfo {pages} {193603} (\bibinfo {year} {2015}{\natexlab{a}})}\BibitemShut
  {NoStop}%
\bibitem [{\citenamefont {Krause}\ \emph {et~al.}(1992)\citenamefont {Krause},
  \citenamefont {Schafer},\ and\ \citenamefont
  {Kulander}}]{PhysRevLett.68.3535}%
  \BibitemOpen
  \bibfield  {author} {\bibinfo {author} {\bibfnamefont {J.~L.}\ \bibnamefont
  {Krause}}, \bibinfo {author} {\bibfnamefont {K.~J.}\ \bibnamefont
  {Schafer}},\ and\ \bibinfo {author} {\bibfnamefont {K.~C.}\ \bibnamefont
  {Kulander}},\ }\bibfield  {title} {\bibinfo {title} {High-order harmonic
  generation from atoms and ions in the high intensity regime},\ }\href
  {https://doi.org/10.1103/PhysRevLett.68.3535} {\bibfield  {journal} {\bibinfo
   {journal} {Phys. Rev. Lett.}\ }\textbf {\bibinfo {volume} {68}},\ \bibinfo
  {pages} {3535} (\bibinfo {year} {1992})}\BibitemShut {NoStop}%
\bibitem [{\citenamefont {Lewenstein}\ \emph {et~al.}(1994)\citenamefont
  {Lewenstein}, \citenamefont {Balcou}, \citenamefont {Ivanov}, \citenamefont
  {L'Huillier},\ and\ \citenamefont {Corkum}}]{PhysRevA.49.2117}%
  \BibitemOpen
  \bibfield  {author} {\bibinfo {author} {\bibfnamefont {M.}~\bibnamefont
  {Lewenstein}}, \bibinfo {author} {\bibfnamefont {P.}~\bibnamefont {Balcou}},
  \bibinfo {author} {\bibfnamefont {M.~Y.}\ \bibnamefont {Ivanov}}, \bibinfo
  {author} {\bibfnamefont {A.}~\bibnamefont {L'Huillier}},\ and\ \bibinfo
  {author} {\bibfnamefont {P.~B.}\ \bibnamefont {Corkum}},\ }\bibfield  {title}
  {\bibinfo {title} {Theory of high-harmonic generation by low-frequency laser
  fields},\ }\href {https://doi.org/10.1103/PhysRevA.49.2117} {\bibfield
  {journal} {\bibinfo  {journal} {Phys. Rev. A}\ }\textbf {\bibinfo {volume}
  {49}},\ \bibinfo {pages} {2117} (\bibinfo {year} {1994})}\BibitemShut
  {NoStop}%
\bibitem [{\citenamefont {Corkum}(1993)}]{PhysRevLett.71.1994}%
  \BibitemOpen
  \bibfield  {author} {\bibinfo {author} {\bibfnamefont {P.~B.}\ \bibnamefont
  {Corkum}},\ }\bibfield  {title} {\bibinfo {title} {Plasma perspective on
  strong field multiphoton ionization},\ }\href
  {https://doi.org/10.1103/PhysRevLett.71.1994} {\bibfield  {journal} {\bibinfo
   {journal} {Phys. Rev. Lett.}\ }\textbf {\bibinfo {volume} {71}},\ \bibinfo
  {pages} {1994} (\bibinfo {year} {1993})}\BibitemShut {NoStop}%
\bibitem [{\citenamefont {Ghimire}\ \emph {et~al.}(2012)\citenamefont
  {Ghimire}, \citenamefont {DiChiara}, \citenamefont {Sistrunk}, \citenamefont
  {Ndabashimiye}, \citenamefont {Szafruga}, \citenamefont {Mohammad},
  \citenamefont {Agostini}, \citenamefont {DiMauro},\ and\ \citenamefont
  {Reis}}]{PhysRevA.85.043836}%
  \BibitemOpen
  \bibfield  {author} {\bibinfo {author} {\bibfnamefont {S.}~\bibnamefont
  {Ghimire}}, \bibinfo {author} {\bibfnamefont {A.~D.}\ \bibnamefont
  {DiChiara}}, \bibinfo {author} {\bibfnamefont {E.}~\bibnamefont {Sistrunk}},
  \bibinfo {author} {\bibfnamefont {G.}~\bibnamefont {Ndabashimiye}}, \bibinfo
  {author} {\bibfnamefont {U.~B.}\ \bibnamefont {Szafruga}}, \bibinfo {author}
  {\bibfnamefont {A.}~\bibnamefont {Mohammad}}, \bibinfo {author}
  {\bibfnamefont {P.}~\bibnamefont {Agostini}}, \bibinfo {author}
  {\bibfnamefont {L.~F.}\ \bibnamefont {DiMauro}},\ and\ \bibinfo {author}
  {\bibfnamefont {D.~A.}\ \bibnamefont {Reis}},\ }\bibfield  {title} {\bibinfo
  {title} {Generation and propagation of high-order harmonics in crystals},\
  }\href {https://doi.org/10.1103/PhysRevA.85.043836} {\bibfield  {journal}
  {\bibinfo  {journal} {Phys. Rev. A}\ }\textbf {\bibinfo {volume} {85}},\
  \bibinfo {pages} {043836} (\bibinfo {year} {2012})}\BibitemShut {NoStop}%
\bibitem [{\citenamefont {Hawkins}\ \emph {et~al.}(2015)\citenamefont
  {Hawkins}, \citenamefont {Ivanov},\ and\ \citenamefont
  {Yakovlev}}]{PhysRevA.91.013405}%
  \BibitemOpen
  \bibfield  {author} {\bibinfo {author} {\bibfnamefont {P.~G.}\ \bibnamefont
  {Hawkins}}, \bibinfo {author} {\bibfnamefont {M.~Y.}\ \bibnamefont
  {Ivanov}},\ and\ \bibinfo {author} {\bibfnamefont {V.~S.}\ \bibnamefont
  {Yakovlev}},\ }\bibfield  {title} {\bibinfo {title} {Effect of multiple
  conduction bands on high-harmonic emission from dielectrics},\ }\href
  {https://doi.org/10.1103/PhysRevA.91.013405} {\bibfield  {journal} {\bibinfo
  {journal} {Phys. Rev. A}\ }\textbf {\bibinfo {volume} {91}},\ \bibinfo
  {pages} {013405} (\bibinfo {year} {2015})}\BibitemShut {NoStop}%
\bibitem [{\citenamefont {Lakhotia}\ \emph {et~al.}(2020)\citenamefont
  {Lakhotia}, \citenamefont {Kim}, \citenamefont {Zhan}, \citenamefont {Hu},
  \citenamefont {Meng},\ and\ \citenamefont {Goulielmakis}}]{Lakhotia2020}%
  \BibitemOpen
  \bibfield  {author} {\bibinfo {author} {\bibfnamefont {H.}~\bibnamefont
  {Lakhotia}}, \bibinfo {author} {\bibfnamefont {H.~Y.}\ \bibnamefont {Kim}},
  \bibinfo {author} {\bibfnamefont {M.}~\bibnamefont {Zhan}}, \bibinfo {author}
  {\bibfnamefont {S.}~\bibnamefont {Hu}}, \bibinfo {author} {\bibfnamefont
  {S.}~\bibnamefont {Meng}},\ and\ \bibinfo {author} {\bibfnamefont
  {E.}~\bibnamefont {Goulielmakis}},\ }\bibfield  {title} {\bibinfo {title}
  {Laser picoscopy of valence electrons in solids},\ }\href
  {https://doi.org/10.1038/s41586-020-2429-z} {\bibfield  {journal} {\bibinfo
  {journal} {Nature}\ }\textbf {\bibinfo {volume} {583}},\ \bibinfo {pages}
  {55} (\bibinfo {year} {2020})}\BibitemShut {NoStop}%
\bibitem [{\citenamefont {Yue}\ and\ \citenamefont
  {Gaarde}(2020{\natexlab{a}})}]{PhysRevA.101.053411}%
  \BibitemOpen
  \bibfield  {author} {\bibinfo {author} {\bibfnamefont {L.}~\bibnamefont
  {Yue}}\ and\ \bibinfo {author} {\bibfnamefont {M.~B.}\ \bibnamefont
  {Gaarde}},\ }\bibfield  {title} {\bibinfo {title} {Structure gauges and laser
  gauges for the semiconductor {{B}}loch equations in high-order harmonic
  generation in solids},\ }\href {https://doi.org/10.1103/PhysRevA.101.053411}
  {\bibfield  {journal} {\bibinfo  {journal} {Phys. Rev. A}\ }\textbf {\bibinfo
  {volume} {101}},\ \bibinfo {pages} {053411} (\bibinfo {year}
  {2020}{\natexlab{a}})}\BibitemShut {NoStop}%
\bibitem [{\citenamefont {Yue}\ and\ \citenamefont
  {Gaarde}(2020{\natexlab{b}})}]{PhysRevLett.124.153204}%
  \BibitemOpen
  \bibfield  {author} {\bibinfo {author} {\bibfnamefont {L.}~\bibnamefont
  {Yue}}\ and\ \bibinfo {author} {\bibfnamefont {M.~B.}\ \bibnamefont
  {Gaarde}},\ }\bibfield  {title} {\bibinfo {title} {Imperfect recollisions in
  high-harmonic generation in solids},\ }\href
  {https://doi.org/10.1103/PhysRevLett.124.153204} {\bibfield  {journal}
  {\bibinfo  {journal} {Phys. Rev. Lett.}\ }\textbf {\bibinfo {volume} {124}},\
  \bibinfo {pages} {153204} (\bibinfo {year} {2020}{\natexlab{b}})}\BibitemShut
  {NoStop}%
\bibitem [{\citenamefont {Tamaya}\ \emph {et~al.}(2016)\citenamefont {Tamaya},
  \citenamefont {Ishikawa}, \citenamefont {Ogawa},\ and\ \citenamefont
  {Tanaka}}]{PhysRevLett.116.016601}%
  \BibitemOpen
  \bibfield  {author} {\bibinfo {author} {\bibfnamefont {T.}~\bibnamefont
  {Tamaya}}, \bibinfo {author} {\bibfnamefont {A.}~\bibnamefont {Ishikawa}},
  \bibinfo {author} {\bibfnamefont {T.}~\bibnamefont {Ogawa}},\ and\ \bibinfo
  {author} {\bibfnamefont {K.}~\bibnamefont {Tanaka}},\ }\bibfield  {title}
  {\bibinfo {title} {Diabatic mechanisms of higher-order harmonic generation in
  solid-state materials under high-intensity electric fields},\ }\href
  {https://doi.org/10.1103/PhysRevLett.116.016601} {\bibfield  {journal}
  {\bibinfo  {journal} {Phys. Rev. Lett.}\ }\textbf {\bibinfo {volume} {116}},\
  \bibinfo {pages} {016601} (\bibinfo {year} {2016})}\BibitemShut {NoStop}%
\bibitem [{\citenamefont {Vampa}\ \emph {et~al.}(2014)\citenamefont {Vampa},
  \citenamefont {McDonald}, \citenamefont {Orlando}, \citenamefont {Klug},
  \citenamefont {Corkum},\ and\ \citenamefont
  {Brabec}}]{PhysRevLett.113.073901}%
  \BibitemOpen
  \bibfield  {author} {\bibinfo {author} {\bibfnamefont {G.}~\bibnamefont
  {Vampa}}, \bibinfo {author} {\bibfnamefont {C.~R.}\ \bibnamefont {McDonald}},
  \bibinfo {author} {\bibfnamefont {G.}~\bibnamefont {Orlando}}, \bibinfo
  {author} {\bibfnamefont {D.~D.}\ \bibnamefont {Klug}}, \bibinfo {author}
  {\bibfnamefont {P.~B.}\ \bibnamefont {Corkum}},\ and\ \bibinfo {author}
  {\bibfnamefont {T.}~\bibnamefont {Brabec}},\ }\bibfield  {title} {\bibinfo
  {title} {Theoretical analysis of high-harmonic generation in solids},\ }\href
  {https://doi.org/10.1103/PhysRevLett.113.073901} {\bibfield  {journal}
  {\bibinfo  {journal} {Phys. Rev. Lett.}\ }\textbf {\bibinfo {volume} {113}},\
  \bibinfo {pages} {073901} (\bibinfo {year} {2014})}\BibitemShut {NoStop}%
\bibitem [{\citenamefont {Higuchi}\ \emph {et~al.}(2014)\citenamefont
  {Higuchi}, \citenamefont {Stockman},\ and\ \citenamefont
  {Hommelhoff}}]{PhysRevLett.113.213901}%
  \BibitemOpen
  \bibfield  {author} {\bibinfo {author} {\bibfnamefont {T.}~\bibnamefont
  {Higuchi}}, \bibinfo {author} {\bibfnamefont {M.~I.}\ \bibnamefont
  {Stockman}},\ and\ \bibinfo {author} {\bibfnamefont {P.}~\bibnamefont
  {Hommelhoff}},\ }\bibfield  {title} {\bibinfo {title} {Strong-field
  perspective on high-harmonic radiation from bulk solids},\ }\href
  {https://doi.org/10.1103/PhysRevLett.113.213901} {\bibfield  {journal}
  {\bibinfo  {journal} {Phys. Rev. Lett.}\ }\textbf {\bibinfo {volume} {113}},\
  \bibinfo {pages} {213901} (\bibinfo {year} {2014})}\BibitemShut {NoStop}%
\bibitem [{\citenamefont {Ikemachi}\ \emph {et~al.}(2017)\citenamefont
  {Ikemachi}, \citenamefont {Shinohara}, \citenamefont {Sato}, \citenamefont
  {Yumoto}, \citenamefont {Kuwata-Gonokami},\ and\ \citenamefont
  {Ishikawa}}]{PhysRevA.95.043416}%
  \BibitemOpen
  \bibfield  {author} {\bibinfo {author} {\bibfnamefont {T.}~\bibnamefont
  {Ikemachi}}, \bibinfo {author} {\bibfnamefont {Y.}~\bibnamefont {Shinohara}},
  \bibinfo {author} {\bibfnamefont {T.}~\bibnamefont {Sato}}, \bibinfo {author}
  {\bibfnamefont {J.}~\bibnamefont {Yumoto}}, \bibinfo {author} {\bibfnamefont
  {M.}~\bibnamefont {Kuwata-Gonokami}},\ and\ \bibinfo {author} {\bibfnamefont
  {K.~L.}\ \bibnamefont {Ishikawa}},\ }\bibfield  {title} {\bibinfo {title}
  {Trajectory analysis of high-order-harmonic generation from periodic
  crystals},\ }\href {https://doi.org/10.1103/PhysRevA.95.043416} {\bibfield
  {journal} {\bibinfo  {journal} {Phys. Rev. A}\ }\textbf {\bibinfo {volume}
  {95}},\ \bibinfo {pages} {043416} (\bibinfo {year} {2017})}\BibitemShut
  {NoStop}%
\bibitem [{\citenamefont {Li}\ \emph {et~al.}(2019)\citenamefont {Li},
  \citenamefont {Lan}, \citenamefont {Zhu}, \citenamefont {Huang},
  \citenamefont {Zhang}, \citenamefont {Lein},\ and\ \citenamefont
  {Lu}}]{PhysRevLett.122.193901}%
  \BibitemOpen
  \bibfield  {author} {\bibinfo {author} {\bibfnamefont {L.}~\bibnamefont
  {Li}}, \bibinfo {author} {\bibfnamefont {P.}~\bibnamefont {Lan}}, \bibinfo
  {author} {\bibfnamefont {X.}~\bibnamefont {Zhu}}, \bibinfo {author}
  {\bibfnamefont {T.}~\bibnamefont {Huang}}, \bibinfo {author} {\bibfnamefont
  {Q.}~\bibnamefont {Zhang}}, \bibinfo {author} {\bibfnamefont
  {M.}~\bibnamefont {Lein}},\ and\ \bibinfo {author} {\bibfnamefont
  {P.}~\bibnamefont {Lu}},\ }\bibfield  {title} {\bibinfo {title}
  {Reciprocal-space-trajectory perspective on high-harmonic generation in
  solids},\ }\href {https://doi.org/10.1103/PhysRevLett.122.193901} {\bibfield
  {journal} {\bibinfo  {journal} {Phys. Rev. Lett.}\ }\textbf {\bibinfo
  {volume} {122}},\ \bibinfo {pages} {193901} (\bibinfo {year}
  {2019})}\BibitemShut {NoStop}%
\bibitem [{\citenamefont {Hohenleutner}\ \emph {et~al.}(2015)\citenamefont
  {Hohenleutner}, \citenamefont {Langer}, \citenamefont {Schubert},
  \citenamefont {Knorr}, \citenamefont {Huttner}, \citenamefont {Koch},
  \citenamefont {Kira},\ and\ \citenamefont {Huber}}]{Hohenleutner2015}%
  \BibitemOpen
  \bibfield  {author} {\bibinfo {author} {\bibfnamefont {M.}~\bibnamefont
  {Hohenleutner}}, \bibinfo {author} {\bibfnamefont {F.}~\bibnamefont
  {Langer}}, \bibinfo {author} {\bibfnamefont {O.}~\bibnamefont {Schubert}},
  \bibinfo {author} {\bibfnamefont {M.}~\bibnamefont {Knorr}}, \bibinfo
  {author} {\bibfnamefont {U.}~\bibnamefont {Huttner}}, \bibinfo {author}
  {\bibfnamefont {S.~W.}\ \bibnamefont {Koch}}, \bibinfo {author}
  {\bibfnamefont {M.}~\bibnamefont {Kira}},\ and\ \bibinfo {author}
  {\bibfnamefont {R.}~\bibnamefont {Huber}},\ }\bibfield  {title} {\bibinfo
  {title} {Real-time observation of interfering crystal electrons in
  high-harmonic generation},\ }\href {https://doi.org/10.1038/nature14652}
  {\bibfield  {journal} {\bibinfo  {journal} {Nature}\ }\textbf {\bibinfo
  {volume} {523}},\ \bibinfo {pages} {572} (\bibinfo {year}
  {2015})}\BibitemShut {NoStop}%
\bibitem [{\citenamefont {Wu}\ \emph {et~al.}(2016)\citenamefont {Wu},
  \citenamefont {Browne}, \citenamefont {Schafer},\ and\ \citenamefont
  {Gaarde}}]{PhysRevA.94.063403}%
  \BibitemOpen
  \bibfield  {author} {\bibinfo {author} {\bibfnamefont {M.}~\bibnamefont
  {Wu}}, \bibinfo {author} {\bibfnamefont {D.~A.}\ \bibnamefont {Browne}},
  \bibinfo {author} {\bibfnamefont {K.~J.}\ \bibnamefont {Schafer}},\ and\
  \bibinfo {author} {\bibfnamefont {M.~B.}\ \bibnamefont {Gaarde}},\ }\bibfield
   {title} {\bibinfo {title} {Multilevel perspective on high-order harmonic
  generation in solids},\ }\href {https://doi.org/10.1103/PhysRevA.94.063403}
  {\bibfield  {journal} {\bibinfo  {journal} {Phys. Rev. A}\ }\textbf {\bibinfo
  {volume} {94}},\ \bibinfo {pages} {063403} (\bibinfo {year}
  {2016})}\BibitemShut {NoStop}%
\bibitem [{\citenamefont {You}\ \emph {et~al.}(2017)\citenamefont {You},
  \citenamefont {Reis},\ and\ \citenamefont {Ghimire}}]{You2017}%
  \BibitemOpen
  \bibfield  {author} {\bibinfo {author} {\bibfnamefont {Y.~S.}\ \bibnamefont
  {You}}, \bibinfo {author} {\bibfnamefont {D.~A.}\ \bibnamefont {Reis}},\ and\
  \bibinfo {author} {\bibfnamefont {S.}~\bibnamefont {Ghimire}},\ }\bibfield
  {title} {\bibinfo {title} {Anisotropic high-harmonic generation in bulk
  crystals},\ }\href {https://doi.org/10.1038/nphys3955} {\bibfield  {journal}
  {\bibinfo  {journal} {Nature Physics}\ }\textbf {\bibinfo {volume} {13}},\
  \bibinfo {pages} {345} (\bibinfo {year} {2017})}\BibitemShut {NoStop}%
\bibitem [{\citenamefont {Golde}\ \emph {et~al.}(2008)\citenamefont {Golde},
  \citenamefont {Meier},\ and\ \citenamefont {Koch}}]{PhysRevB.77.075330}%
  \BibitemOpen
  \bibfield  {author} {\bibinfo {author} {\bibfnamefont {D.}~\bibnamefont
  {Golde}}, \bibinfo {author} {\bibfnamefont {T.}~\bibnamefont {Meier}},\ and\
  \bibinfo {author} {\bibfnamefont {S.~W.}\ \bibnamefont {Koch}},\ }\bibfield
  {title} {\bibinfo {title} {High harmonics generated in semiconductor
  nanostructures by the coupled dynamics of optical inter- and intraband
  excitations},\ }\href {https://doi.org/10.1103/PhysRevB.77.075330} {\bibfield
   {journal} {\bibinfo  {journal} {Phys. Rev. B}\ }\textbf {\bibinfo {volume}
  {77}},\ \bibinfo {pages} {075330} (\bibinfo {year} {2008})}\BibitemShut
  {NoStop}%
\bibitem [{\citenamefont {Vampa}\ \emph
  {et~al.}(2015{\natexlab{b}})\citenamefont {Vampa}, \citenamefont {McDonald},
  \citenamefont {Orlando}, \citenamefont {Corkum},\ and\ \citenamefont
  {Brabec}}]{PhysRevB.91.064302}%
  \BibitemOpen
  \bibfield  {author} {\bibinfo {author} {\bibfnamefont {G.}~\bibnamefont
  {Vampa}}, \bibinfo {author} {\bibfnamefont {C.~R.}\ \bibnamefont {McDonald}},
  \bibinfo {author} {\bibfnamefont {G.}~\bibnamefont {Orlando}}, \bibinfo
  {author} {\bibfnamefont {P.~B.}\ \bibnamefont {Corkum}},\ and\ \bibinfo
  {author} {\bibfnamefont {T.}~\bibnamefont {Brabec}},\ }\bibfield  {title}
  {\bibinfo {title} {Semiclassical analysis of high harmonic generation in bulk
  crystals},\ }\href {https://doi.org/10.1103/PhysRevB.91.064302} {\bibfield
  {journal} {\bibinfo  {journal} {Phys. Rev. B}\ }\textbf {\bibinfo {volume}
  {91}},\ \bibinfo {pages} {064302} (\bibinfo {year}
  {2015}{\natexlab{b}})}\BibitemShut {NoStop}%
\bibitem [{\citenamefont {Ghimire}\ \emph {et~al.}(2011)\citenamefont
  {Ghimire}, \citenamefont {DiChiara}, \citenamefont {Sistrunk}, \citenamefont
  {Agostini}, \citenamefont {DiMauro},\ and\ \citenamefont
  {Reis}}]{Ghimire2011}%
  \BibitemOpen
  \bibfield  {author} {\bibinfo {author} {\bibfnamefont {S.}~\bibnamefont
  {Ghimire}}, \bibinfo {author} {\bibfnamefont {A.~D.}\ \bibnamefont
  {DiChiara}}, \bibinfo {author} {\bibfnamefont {E.}~\bibnamefont {Sistrunk}},
  \bibinfo {author} {\bibfnamefont {P.}~\bibnamefont {Agostini}}, \bibinfo
  {author} {\bibfnamefont {L.~F.}\ \bibnamefont {DiMauro}},\ and\ \bibinfo
  {author} {\bibfnamefont {D.~A.}\ \bibnamefont {Reis}},\ }\bibfield  {title}
  {\bibinfo {title} {Observation of high-order harmonic generation in a bulk
  crystal},\ }\href {https://doi.org/10.1038/nphys1847} {\bibfield  {journal}
  {\bibinfo  {journal} {Nature Physics}\ }\textbf {\bibinfo {volume} {7}},\
  \bibinfo {pages} {138} (\bibinfo {year} {2011})}\BibitemShut {NoStop}%
\bibitem [{\citenamefont {Kaneshima}\ \emph {et~al.}(2018)\citenamefont
  {Kaneshima}, \citenamefont {Shinohara}, \citenamefont {Takeuchi},
  \citenamefont {Ishii}, \citenamefont {Imasaka}, \citenamefont {Kaji},
  \citenamefont {Ashihara}, \citenamefont {Ishikawa},\ and\ \citenamefont
  {Itatani}}]{PhysRevLett.120.243903}%
  \BibitemOpen
  \bibfield  {author} {\bibinfo {author} {\bibfnamefont {K.}~\bibnamefont
  {Kaneshima}}, \bibinfo {author} {\bibfnamefont {Y.}~\bibnamefont
  {Shinohara}}, \bibinfo {author} {\bibfnamefont {K.}~\bibnamefont {Takeuchi}},
  \bibinfo {author} {\bibfnamefont {N.}~\bibnamefont {Ishii}}, \bibinfo
  {author} {\bibfnamefont {K.}~\bibnamefont {Imasaka}}, \bibinfo {author}
  {\bibfnamefont {T.}~\bibnamefont {Kaji}}, \bibinfo {author} {\bibfnamefont
  {S.}~\bibnamefont {Ashihara}}, \bibinfo {author} {\bibfnamefont {K.~L.}\
  \bibnamefont {Ishikawa}},\ and\ \bibinfo {author} {\bibfnamefont
  {J.}~\bibnamefont {Itatani}},\ }\bibfield  {title} {\bibinfo {title}
  {Polarization-resolved study of high harmonics from bulk semiconductors},\
  }\href {https://doi.org/10.1103/PhysRevLett.120.243903} {\bibfield  {journal}
  {\bibinfo  {journal} {Phys. Rev. Lett.}\ }\textbf {\bibinfo {volume} {120}},\
  \bibinfo {pages} {243903} (\bibinfo {year} {2018})}\BibitemShut {NoStop}%
\bibitem [{\citenamefont {Klemke}\ \emph {et~al.}(2020)\citenamefont {Klemke},
  \citenamefont {M\"ucke}, \citenamefont {Rubio}, \citenamefont {K\"artner},\
  and\ \citenamefont {Tancogne-Dejean}}]{PhysRevB.102.104308}%
  \BibitemOpen
  \bibfield  {author} {\bibinfo {author} {\bibfnamefont {N.}~\bibnamefont
  {Klemke}}, \bibinfo {author} {\bibfnamefont {O.~D.}\ \bibnamefont {M\"ucke}},
  \bibinfo {author} {\bibfnamefont {A.}~\bibnamefont {Rubio}}, \bibinfo
  {author} {\bibfnamefont {F.~X.}\ \bibnamefont {K\"artner}},\ and\ \bibinfo
  {author} {\bibfnamefont {N.}~\bibnamefont {Tancogne-Dejean}},\ }\bibfield
  {title} {\bibinfo {title} {Role of intraband dynamics in the generation of
  circularly polarized high harmonics from solids},\ }\href
  {https://doi.org/10.1103/PhysRevB.102.104308} {\bibfield  {journal} {\bibinfo
   {journal} {Phys. Rev. B}\ }\textbf {\bibinfo {volume} {102}},\ \bibinfo
  {pages} {104308} (\bibinfo {year} {2020})}\BibitemShut {NoStop}%
\bibitem [{\citenamefont {Lanin}\ \emph {et~al.}(2017)\citenamefont {Lanin},
  \citenamefont {Stepanov}, \citenamefont {Fedotov},\ and\ \citenamefont
  {Zheltikov}}]{Lanin:17}%
  \BibitemOpen
  \bibfield  {author} {\bibinfo {author} {\bibfnamefont {A.~A.}\ \bibnamefont
  {Lanin}}, \bibinfo {author} {\bibfnamefont {E.~A.}\ \bibnamefont {Stepanov}},
  \bibinfo {author} {\bibfnamefont {A.~B.}\ \bibnamefont {Fedotov}},\ and\
  \bibinfo {author} {\bibfnamefont {A.~M.}\ \bibnamefont {Zheltikov}},\
  }\bibfield  {title} {\bibinfo {title} {Mapping the electron band structure by
  intraband high-harmonic generation in solids},\ }\href
  {https://doi.org/10.1364/OPTICA.4.000516} {\bibfield  {journal} {\bibinfo
  {journal} {Optica}\ }\textbf {\bibinfo {volume} {4}},\ \bibinfo {pages} {516}
  (\bibinfo {year} {2017})}\BibitemShut {NoStop}%
\bibitem [{\citenamefont {Lanin}\ \emph {et~al.}(2019)\citenamefont {Lanin},
  \citenamefont {Stepanov}, \citenamefont {Mitrofanov}, \citenamefont
  {Sidorov-Biryukov}, \citenamefont {Fedotov},\ and\ \citenamefont
  {Zheltikov}}]{Lanin:19}%
  \BibitemOpen
  \bibfield  {author} {\bibinfo {author} {\bibfnamefont {A.~A.}\ \bibnamefont
  {Lanin}}, \bibinfo {author} {\bibfnamefont {E.~A.}\ \bibnamefont {Stepanov}},
  \bibinfo {author} {\bibfnamefont {A.~V.}\ \bibnamefont {Mitrofanov}},
  \bibinfo {author} {\bibfnamefont {D.~A.}\ \bibnamefont {Sidorov-Biryukov}},
  \bibinfo {author} {\bibfnamefont {A.~B.}\ \bibnamefont {Fedotov}},\ and\
  \bibinfo {author} {\bibfnamefont {A.~M.}\ \bibnamefont {Zheltikov}},\
  }\bibfield  {title} {\bibinfo {title} {High-order harmonic analysis of
  anisotropic petahertz photocurrents in solids},\ }\href
  {https://doi.org/10.1364/OL.44.001888} {\bibfield  {journal} {\bibinfo
  {journal} {Opt. Lett.}\ }\textbf {\bibinfo {volume} {44}},\ \bibinfo {pages}
  {1888} (\bibinfo {year} {2019})}\BibitemShut {NoStop}%
\bibitem [{\citenamefont {Liu}\ \emph {et~al.}(2017)\citenamefont {Liu},
  \citenamefont {Li}, \citenamefont {You}, \citenamefont {Ghimire},
  \citenamefont {Heinz},\ and\ \citenamefont {Reis}}]{Liu2017}%
  \BibitemOpen
  \bibfield  {author} {\bibinfo {author} {\bibfnamefont {H.}~\bibnamefont
  {Liu}}, \bibinfo {author} {\bibfnamefont {Y.}~\bibnamefont {Li}}, \bibinfo
  {author} {\bibfnamefont {Y.~S.}\ \bibnamefont {You}}, \bibinfo {author}
  {\bibfnamefont {S.}~\bibnamefont {Ghimire}}, \bibinfo {author} {\bibfnamefont
  {T.~F.}\ \bibnamefont {Heinz}},\ and\ \bibinfo {author} {\bibfnamefont
  {D.~A.}\ \bibnamefont {Reis}},\ }\bibfield  {title} {\bibinfo {title}
  {High-harmonic generation from an atomically thin semiconductor},\ }\href
  {https://doi.org/10.1038/nphys3946} {\bibfield  {journal} {\bibinfo
  {journal} {Nature Physics}\ }\textbf {\bibinfo {volume} {13}},\ \bibinfo
  {pages} {262} (\bibinfo {year} {2017})}\BibitemShut {NoStop}%
\bibitem [{\citenamefont {Luu}\ \emph {et~al.}(2015)\citenamefont {Luu},
  \citenamefont {Garg}, \citenamefont {Kruchinin}, \citenamefont {Moulet},
  \citenamefont {Hassan},\ and\ \citenamefont {Goulielmakis}}]{Luu2015}%
  \BibitemOpen
  \bibfield  {author} {\bibinfo {author} {\bibfnamefont {T.~T.}\ \bibnamefont
  {Luu}}, \bibinfo {author} {\bibfnamefont {M.}~\bibnamefont {Garg}}, \bibinfo
  {author} {\bibfnamefont {S.~Y.}\ \bibnamefont {Kruchinin}}, \bibinfo {author}
  {\bibfnamefont {A.}~\bibnamefont {Moulet}}, \bibinfo {author} {\bibfnamefont
  {M.~T.}\ \bibnamefont {Hassan}},\ and\ \bibinfo {author} {\bibfnamefont
  {E.}~\bibnamefont {Goulielmakis}},\ }\bibfield  {title} {\bibinfo {title}
  {Extreme ultraviolet high-harmonic spectroscopy of solids},\ }\href
  {https://doi.org/10.1038/nature14456} {\bibfield  {journal} {\bibinfo
  {journal} {Nature}\ }\textbf {\bibinfo {volume} {521}},\ \bibinfo {pages}
  {498} (\bibinfo {year} {2015})}\BibitemShut {NoStop}%
\bibitem [{\citenamefont {Luu}\ and\ \citenamefont
  {W{\"o}rner}(2018)}]{Luu2018}%
  \BibitemOpen
  \bibfield  {author} {\bibinfo {author} {\bibfnamefont {T.~T.}\ \bibnamefont
  {Luu}}\ and\ \bibinfo {author} {\bibfnamefont {H.~J.}\ \bibnamefont
  {W{\"o}rner}},\ }\bibfield  {title} {\bibinfo {title} {Measurement of the
  {{B}}erry curvature of solids using high-harmonic spectroscopy},\ }\href
  {https://doi.org/10.1038/s41467-018-03397-4} {\bibfield  {journal} {\bibinfo
  {journal} {Nature Communications}\ }\textbf {\bibinfo {volume} {9}},\
  \bibinfo {pages} {916} (\bibinfo {year} {2018})}\BibitemShut {NoStop}%
\bibitem [{\citenamefont {Ludwig}\ \emph {et~al.}(2014)\citenamefont {Ludwig},
  \citenamefont {Maurer}, \citenamefont {Mayer}, \citenamefont {Phillips},
  \citenamefont {Gallmann},\ and\ \citenamefont
  {Keller}}]{PhysRevLett.113.243001}%
  \BibitemOpen
  \bibfield  {author} {\bibinfo {author} {\bibfnamefont {A.}~\bibnamefont
  {Ludwig}}, \bibinfo {author} {\bibfnamefont {J.}~\bibnamefont {Maurer}},
  \bibinfo {author} {\bibfnamefont {B.~W.}\ \bibnamefont {Mayer}}, \bibinfo
  {author} {\bibfnamefont {C.~R.}\ \bibnamefont {Phillips}}, \bibinfo {author}
  {\bibfnamefont {L.}~\bibnamefont {Gallmann}},\ and\ \bibinfo {author}
  {\bibfnamefont {U.}~\bibnamefont {Keller}},\ }\bibfield  {title} {\bibinfo
  {title} {Breakdown of the dipole approximation in strong-field ionization},\
  }\href {https://doi.org/10.1103/PhysRevLett.113.243001} {\bibfield  {journal}
  {\bibinfo  {journal} {Phys. Rev. Lett.}\ }\textbf {\bibinfo {volume} {113}},\
  \bibinfo {pages} {243001} (\bibinfo {year} {2014})}\BibitemShut {NoStop}%
\bibitem [{\citenamefont {Willenberg}\ \emph {et~al.}(2019)\citenamefont
  {Willenberg}, \citenamefont {Maurer}, \citenamefont {Mayer},\ and\
  \citenamefont {Keller}}]{Willenberg2019}%
  \BibitemOpen
  \bibfield  {author} {\bibinfo {author} {\bibfnamefont {B.}~\bibnamefont
  {Willenberg}}, \bibinfo {author} {\bibfnamefont {J.}~\bibnamefont {Maurer}},
  \bibinfo {author} {\bibfnamefont {B.~W.}\ \bibnamefont {Mayer}},\ and\
  \bibinfo {author} {\bibfnamefont {U.}~\bibnamefont {Keller}},\ }\bibfield
  {title} {\bibinfo {title} {Sub-cycle time resolution of multi-photon momentum
  transfer in strong-field ionization},\ }\href
  {https://doi.org/10.1038/s41467-019-13409-6} {\bibfield  {journal} {\bibinfo
  {journal} {Nature Communications}\ }\textbf {\bibinfo {volume} {10}},\
  \bibinfo {pages} {5548} (\bibinfo {year} {2019})}\BibitemShut {NoStop}%
\bibitem [{\citenamefont {Haram}\ \emph {et~al.}(2019)\citenamefont {Haram},
  \citenamefont {Ivanov}, \citenamefont {Xu}, \citenamefont {Kim},
  \citenamefont {Atia-tul Noor}, \citenamefont {Sainadh}, \citenamefont
  {Glover}, \citenamefont {Chetty}, \citenamefont {Litvinyuk},\ and\
  \citenamefont {Sang}}]{PhysRevLett.123.093201}%
  \BibitemOpen
  \bibfield  {author} {\bibinfo {author} {\bibfnamefont {N.}~\bibnamefont
  {Haram}}, \bibinfo {author} {\bibfnamefont {I.}~\bibnamefont {Ivanov}},
  \bibinfo {author} {\bibfnamefont {H.}~\bibnamefont {Xu}}, \bibinfo {author}
  {\bibfnamefont {K.~T.}\ \bibnamefont {Kim}}, \bibinfo {author} {\bibfnamefont
  {A.}~\bibnamefont {Atia-tul Noor}}, \bibinfo {author} {\bibfnamefont {U.~S.}\
  \bibnamefont {Sainadh}}, \bibinfo {author} {\bibfnamefont {R.~D.}\
  \bibnamefont {Glover}}, \bibinfo {author} {\bibfnamefont {D.}~\bibnamefont
  {Chetty}}, \bibinfo {author} {\bibfnamefont {I.~V.}\ \bibnamefont
  {Litvinyuk}},\ and\ \bibinfo {author} {\bibfnamefont {R.~T.}\ \bibnamefont
  {Sang}},\ }\bibfield  {title} {\bibinfo {title} {Relativistic nondipole
  effects in strong-field atomic ionization at moderate intensities},\ }\href
  {https://doi.org/10.1103/PhysRevLett.123.093201} {\bibfield  {journal}
  {\bibinfo  {journal} {Phys. Rev. Lett.}\ }\textbf {\bibinfo {volume} {123}},\
  \bibinfo {pages} {093201} (\bibinfo {year} {2019})}\BibitemShut {NoStop}%
\bibitem [{\citenamefont {Hartung}\ \emph {et~al.}(2021)\citenamefont
  {Hartung}, \citenamefont {Brennecke}, \citenamefont {Lin}, \citenamefont
  {Trabert}, \citenamefont {Fehre}, \citenamefont {Rist}, \citenamefont
  {Sch\"offler}, \citenamefont {Jahnke}, \citenamefont {Schmidt}, \citenamefont
  {Kunitski}, \citenamefont {Lein}, \citenamefont {D\"orner},\ and\
  \citenamefont {Eckart}}]{PhysRevLett.126.053202}%
  \BibitemOpen
  \bibfield  {author} {\bibinfo {author} {\bibfnamefont {A.}~\bibnamefont
  {Hartung}}, \bibinfo {author} {\bibfnamefont {S.}~\bibnamefont {Brennecke}},
  \bibinfo {author} {\bibfnamefont {K.}~\bibnamefont {Lin}}, \bibinfo {author}
  {\bibfnamefont {D.}~\bibnamefont {Trabert}}, \bibinfo {author} {\bibfnamefont
  {K.}~\bibnamefont {Fehre}}, \bibinfo {author} {\bibfnamefont
  {J.}~\bibnamefont {Rist}}, \bibinfo {author} {\bibfnamefont {M.~S.}\
  \bibnamefont {Sch\"offler}}, \bibinfo {author} {\bibfnamefont
  {T.}~\bibnamefont {Jahnke}}, \bibinfo {author} {\bibfnamefont {L.~P.~H.}\
  \bibnamefont {Schmidt}}, \bibinfo {author} {\bibfnamefont {M.}~\bibnamefont
  {Kunitski}}, \bibinfo {author} {\bibfnamefont {M.}~\bibnamefont {Lein}},
  \bibinfo {author} {\bibfnamefont {R.}~\bibnamefont {D\"orner}},\ and\
  \bibinfo {author} {\bibfnamefont {S.}~\bibnamefont {Eckart}},\ }\bibfield
  {title} {\bibinfo {title} {Electric nondipole effect in strong-field
  ionization},\ }\href {https://doi.org/10.1103/PhysRevLett.126.053202}
  {\bibfield  {journal} {\bibinfo  {journal} {Phys. Rev. Lett.}\ }\textbf
  {\bibinfo {volume} {126}},\ \bibinfo {pages} {053202} (\bibinfo {year}
  {2021})}\BibitemShut {NoStop}%
\bibitem [{\citenamefont {Smeenk}\ \emph {et~al.}(2011)\citenamefont {Smeenk},
  \citenamefont {Arissian}, \citenamefont {Zhou}, \citenamefont {Mysyrowicz},
  \citenamefont {Villeneuve}, \citenamefont {Staudte},\ and\ \citenamefont
  {Corkum}}]{PhysRevLett.106.193002}%
  \BibitemOpen
  \bibfield  {author} {\bibinfo {author} {\bibfnamefont {C.~T.~L.}\
  \bibnamefont {Smeenk}}, \bibinfo {author} {\bibfnamefont {L.}~\bibnamefont
  {Arissian}}, \bibinfo {author} {\bibfnamefont {B.}~\bibnamefont {Zhou}},
  \bibinfo {author} {\bibfnamefont {A.}~\bibnamefont {Mysyrowicz}}, \bibinfo
  {author} {\bibfnamefont {D.~M.}\ \bibnamefont {Villeneuve}}, \bibinfo
  {author} {\bibfnamefont {A.}~\bibnamefont {Staudte}},\ and\ \bibinfo {author}
  {\bibfnamefont {P.~B.}\ \bibnamefont {Corkum}},\ }\bibfield  {title}
  {\bibinfo {title} {Partitioning of the linear photon momentum in multiphoton
  ionization},\ }\href {https://doi.org/10.1103/PhysRevLett.106.193002}
  {\bibfield  {journal} {\bibinfo  {journal} {Phys. Rev. Lett.}\ }\textbf
  {\bibinfo {volume} {106}},\ \bibinfo {pages} {193002} (\bibinfo {year}
  {2011})}\BibitemShut {NoStop}%
\bibitem [{\citenamefont {Hartung}\ \emph {et~al.}(2019)\citenamefont
  {Hartung}, \citenamefont {Eckart}, \citenamefont {Brennecke}, \citenamefont
  {Rist}, \citenamefont {Trabert}, \citenamefont {Fehre}, \citenamefont
  {Richter}, \citenamefont {Sann}, \citenamefont {Zeller}, \citenamefont
  {Henrichs}, \citenamefont {Kastirke}, \citenamefont {Hoehl}, \citenamefont
  {Kalinin}, \citenamefont {Sch{\"o}ffler}, \citenamefont {Jahnke},
  \citenamefont {Schmidt}, \citenamefont {Lein}, \citenamefont {Kunitski},\
  and\ \citenamefont {D{\"o}rner}}]{Hartung2019}%
  \BibitemOpen
  \bibfield  {author} {\bibinfo {author} {\bibfnamefont {A.}~\bibnamefont
  {Hartung}}, \bibinfo {author} {\bibfnamefont {S.}~\bibnamefont {Eckart}},
  \bibinfo {author} {\bibfnamefont {S.}~\bibnamefont {Brennecke}}, \bibinfo
  {author} {\bibfnamefont {J.}~\bibnamefont {Rist}}, \bibinfo {author}
  {\bibfnamefont {D.}~\bibnamefont {Trabert}}, \bibinfo {author} {\bibfnamefont
  {K.}~\bibnamefont {Fehre}}, \bibinfo {author} {\bibfnamefont
  {M.}~\bibnamefont {Richter}}, \bibinfo {author} {\bibfnamefont
  {H.}~\bibnamefont {Sann}}, \bibinfo {author} {\bibfnamefont {S.}~\bibnamefont
  {Zeller}}, \bibinfo {author} {\bibfnamefont {K.}~\bibnamefont {Henrichs}},
  \bibinfo {author} {\bibfnamefont {G.}~\bibnamefont {Kastirke}}, \bibinfo
  {author} {\bibfnamefont {J.}~\bibnamefont {Hoehl}}, \bibinfo {author}
  {\bibfnamefont {A.}~\bibnamefont {Kalinin}}, \bibinfo {author} {\bibfnamefont
  {M.~S.}\ \bibnamefont {Sch{\"o}ffler}}, \bibinfo {author} {\bibfnamefont
  {T.}~\bibnamefont {Jahnke}}, \bibinfo {author} {\bibfnamefont {L.~P.~H.}\
  \bibnamefont {Schmidt}}, \bibinfo {author} {\bibfnamefont {M.}~\bibnamefont
  {Lein}}, \bibinfo {author} {\bibfnamefont {M.}~\bibnamefont {Kunitski}},\
  and\ \bibinfo {author} {\bibfnamefont {R.}~\bibnamefont {D{\"o}rner}},\
  }\bibfield  {title} {\bibinfo {title} {Magnetic fields alter strong-field
  ionization},\ }\href {https://doi.org/10.1038/s41567-019-0653-y} {\bibfield
  {journal} {\bibinfo  {journal} {Nature Physics}\ }\textbf {\bibinfo {volume}
  {15}},\ \bibinfo {pages} {1222} (\bibinfo {year} {2019})}\BibitemShut
  {NoStop}%
\bibitem [{\citenamefont {Wang}\ \emph {et~al.}(2020)\citenamefont {Wang},
  \citenamefont {Chen}, \citenamefont {Liang},\ and\ \citenamefont
  {Peng}}]{Wang_2020}%
  \BibitemOpen
  \bibfield  {author} {\bibinfo {author} {\bibfnamefont {M.-X.}\ \bibnamefont
  {Wang}}, \bibinfo {author} {\bibfnamefont {S.-G.}\ \bibnamefont {Chen}},
  \bibinfo {author} {\bibfnamefont {H.}~\bibnamefont {Liang}},\ and\ \bibinfo
  {author} {\bibfnamefont {L.-Y.}\ \bibnamefont {Peng}},\ }\bibfield  {title}
  {\bibinfo {title} {Review on non-dipole effects in ionization and harmonic
  generation of atoms and molecules},\ }\href
  {https://doi.org/10.1088/1674-1056/ab5c10} {\bibfield  {journal} {\bibinfo
  {journal} {Chinese Physics B}\ }\textbf {\bibinfo {volume} {29}},\ \bibinfo
  {pages} {013302} (\bibinfo {year} {2020})}\BibitemShut {NoStop}%
\bibitem [{\citenamefont {Haram}\ \emph {et~al.}(2020)\citenamefont {Haram},
  \citenamefont {Sang},\ and\ \citenamefont {Litvinyuk}}]{Haram_2020}%
  \BibitemOpen
  \bibfield  {author} {\bibinfo {author} {\bibfnamefont {N.}~\bibnamefont
  {Haram}}, \bibinfo {author} {\bibfnamefont {R.~T.}\ \bibnamefont {Sang}},\
  and\ \bibinfo {author} {\bibfnamefont {I.~V.}\ \bibnamefont {Litvinyuk}},\
  }\bibfield  {title} {\bibinfo {title} {Transverse electron momentum
  distributions in strong-field ionization: nondipole and {{C}}oulomb focusing
  effects},\ }\href {https://doi.org/10.1088/1361-6455/ab9272} {\bibfield
  {journal} {\bibinfo  {journal} {J. Phys. B}\ }\textbf {\bibinfo {volume}
  {53}},\ \bibinfo {pages} {154005} (\bibinfo {year} {2020})}\BibitemShut
  {NoStop}%
\bibitem [{\citenamefont {Maurer}\ and\ \citenamefont
  {Keller}(2021)}]{Maurer_2021}%
  \BibitemOpen
  \bibfield  {author} {\bibinfo {author} {\bibfnamefont {J.}~\bibnamefont
  {Maurer}}\ and\ \bibinfo {author} {\bibfnamefont {U.}~\bibnamefont
  {Keller}},\ }\bibfield  {title} {\bibinfo {title} {Ionization in intense
  laser fields beyond the electric dipole approximation: concepts, methods,
  achievements and future directions},\ }\href
  {https://doi.org/10.1088/1361-6455/abf731} {\bibfield  {journal} {\bibinfo
  {journal} {J. Phys. B}\ }\textbf {\bibinfo {volume} {54}},\ \bibinfo {pages}
  {094001} (\bibinfo {year} {2021})}\BibitemShut {NoStop}%
\bibitem [{\citenamefont {Gorlach}\ \emph {et~al.}(2020)\citenamefont
  {Gorlach}, \citenamefont {Neufeld}, \citenamefont {Rivera}, \citenamefont
  {Cohen},\ and\ \citenamefont {Kaminer}}]{Gorlach2020}%
  \BibitemOpen
  \bibfield  {author} {\bibinfo {author} {\bibfnamefont {A.}~\bibnamefont
  {Gorlach}}, \bibinfo {author} {\bibfnamefont {O.}~\bibnamefont {Neufeld}},
  \bibinfo {author} {\bibfnamefont {N.}~\bibnamefont {Rivera}}, \bibinfo
  {author} {\bibfnamefont {O.}~\bibnamefont {Cohen}},\ and\ \bibinfo {author}
  {\bibfnamefont {I.}~\bibnamefont {Kaminer}},\ }\bibfield  {title} {\bibinfo
  {title} {The quantum-optical nature of high harmonic generation},\ }\href
  {https://doi.org/10.1038/s41467-020-18218-w} {\bibfield  {journal} {\bibinfo
  {journal} {Nature Communications}\ }\textbf {\bibinfo {volume} {11}},\
  \bibinfo {pages} {4598} (\bibinfo {year} {2020})}\BibitemShut {NoStop}%
\bibitem [{\citenamefont {Silva}\ \emph {et~al.}(2019)\citenamefont {Silva},
  \citenamefont {Jim{\'e}nez-Gal{\'a}n}, \citenamefont {Amorim}, \citenamefont
  {Smirnova},\ and\ \citenamefont {Ivanov}}]{Silva2019}%
  \BibitemOpen
  \bibfield  {author} {\bibinfo {author} {\bibfnamefont {R.~E.~F.}\
  \bibnamefont {Silva}}, \bibinfo {author} {\bibfnamefont {{\'A}.}~\bibnamefont
  {Jim{\'e}nez-Gal{\'a}n}}, \bibinfo {author} {\bibfnamefont {B.}~\bibnamefont
  {Amorim}}, \bibinfo {author} {\bibfnamefont {O.}~\bibnamefont {Smirnova}},\
  and\ \bibinfo {author} {\bibfnamefont {M.}~\bibnamefont {Ivanov}},\
  }\bibfield  {title} {\bibinfo {title} {Topological strong-field physics on
  sub-laser-cycle timescale},\ }\href
  {https://doi.org/10.1038/s41566-019-0516-1} {\bibfield  {journal} {\bibinfo
  {journal} {Nature Photonics}\ }\textbf {\bibinfo {volume} {13}},\ \bibinfo
  {pages} {849} (\bibinfo {year} {2019})}\BibitemShut {NoStop}%
\bibitem [{\citenamefont {Bugacov}\ \emph {et~al.}(1993)\citenamefont
  {Bugacov}, \citenamefont {Pont},\ and\ \citenamefont
  {Shakeshaft}}]{PhysRevA.48.R4027}%
  \BibitemOpen
  \bibfield  {author} {\bibinfo {author} {\bibfnamefont {A.}~\bibnamefont
  {Bugacov}}, \bibinfo {author} {\bibfnamefont {M.}~\bibnamefont {Pont}},\ and\
  \bibinfo {author} {\bibfnamefont {R.}~\bibnamefont {Shakeshaft}},\ }\bibfield
   {title} {\bibinfo {title} {Possibility of breakdown of atomic stabilization
  in an intense high-frequency field},\ }\href
  {https://doi.org/10.1103/PhysRevA.48.R4027} {\bibfield  {journal} {\bibinfo
  {journal} {Phys. Rev. A}\ }\textbf {\bibinfo {volume} {48}},\ \bibinfo
  {pages} {R4027} (\bibinfo {year} {1993})}\BibitemShut {NoStop}%
\bibitem [{\citenamefont {V\'azquez~de Aldana}\ \emph
  {et~al.}(2001)\citenamefont {V\'azquez~de Aldana}, \citenamefont {Kylstra},
  \citenamefont {Roso}, \citenamefont {Knight}, \citenamefont {Patel},\ and\
  \citenamefont {Worthington}}]{PhysRevA.64.013411}%
  \BibitemOpen
  \bibfield  {author} {\bibinfo {author} {\bibfnamefont {J.~R.}\ \bibnamefont
  {V\'azquez~de Aldana}}, \bibinfo {author} {\bibfnamefont {N.~J.}\
  \bibnamefont {Kylstra}}, \bibinfo {author} {\bibfnamefont {L.}~\bibnamefont
  {Roso}}, \bibinfo {author} {\bibfnamefont {P.~L.}\ \bibnamefont {Knight}},
  \bibinfo {author} {\bibfnamefont {A.}~\bibnamefont {Patel}},\ and\ \bibinfo
  {author} {\bibfnamefont {R.~A.}\ \bibnamefont {Worthington}},\ }\bibfield
  {title} {\bibinfo {title} {Atoms interacting with intense, high-frequency
  laser pulses: Effect of the magnetic-field component on atomic
  stabilization},\ }\href {https://doi.org/10.1103/PhysRevA.64.013411}
  {\bibfield  {journal} {\bibinfo  {journal} {Phys. Rev. A}\ }\textbf {\bibinfo
  {volume} {64}},\ \bibinfo {pages} {013411} (\bibinfo {year}
  {2001})}\BibitemShut {NoStop}%
\bibitem [{\citenamefont {Jensen}\ \emph {et~al.}(2020)\citenamefont {Jensen},
  \citenamefont {Lund},\ and\ \citenamefont {Madsen}}]{PhysRevA.101.043408}%
  \BibitemOpen
  \bibfield  {author} {\bibinfo {author} {\bibfnamefont {S.~V.~B.}\
  \bibnamefont {Jensen}}, \bibinfo {author} {\bibfnamefont {M.~M.}\
  \bibnamefont {Lund}},\ and\ \bibinfo {author} {\bibfnamefont {L.~B.}\
  \bibnamefont {Madsen}},\ }\bibfield  {title} {\bibinfo {title} {Nondipole
  strong-field-approximation {{H}}amiltonian},\ }\href
  {https://doi.org/10.1103/PhysRevA.101.043408} {\bibfield  {journal} {\bibinfo
   {journal} {Phys. Rev. A}\ }\textbf {\bibinfo {volume} {101}},\ \bibinfo
  {pages} {043408} (\bibinfo {year} {2020})}\BibitemShut {NoStop}%
\bibitem [{SM()}]{SM}%
  \BibitemOpen
  \bibfield  {title} {\bibinfo {title} {See {{S}}upplemental {{M}}aterial,
  which includes {{R}}efs. [51,52]},\ }\href@noop {} {\bibinfo  {journal} {for
  a discussion and comparision to HHG in gases and an analytical solution to
  the nondipole strong-field approximation equations of motion}\ }\BibitemShut
  {NoStop}%
\bibitem [{\citenamefont {Gaarde}\ \emph {et~al.}(2008)\citenamefont {Gaarde},
  \citenamefont {Tate},\ and\ \citenamefont {Schafer}}]{Gaarde2008}%
  \BibitemOpen
\bibfield  {journal} {  }\bibfield  {author} {\bibinfo {author} {\bibfnamefont
  {M.~B.}\ \bibnamefont {Gaarde}}, \bibinfo {author} {\bibfnamefont {J.~L.}\
  \bibnamefont {Tate}},\ and\ \bibinfo {author} {\bibfnamefont {K.~J.}\
  \bibnamefont {Schafer}},\ }\bibfield  {title} {\bibinfo {title} {Macroscopic
  aspects of attosecond pulse generation},\ }\href
  {https://doi.org/10.1088/0953-4075/41/13/132001} {\bibfield  {journal}
  {\bibinfo  {journal} {J. Phys. B}\ }\textbf {\bibinfo {volume} {41}},\
  \bibinfo {pages} {132001} (\bibinfo {year} {2008})}\BibitemShut {NoStop}%
\bibitem [{\citenamefont {Bellini}\ \emph {et~al.}(1998)\citenamefont
  {Bellini}, \citenamefont {Lyng\aa{}}, \citenamefont {Tozzi}, \citenamefont
  {Gaarde}, \citenamefont {H\"ansch}, \citenamefont {L'Huillier},\ and\
  \citenamefont {Wahlstr\"om}}]{PhysRevLett.81.297}%
  \BibitemOpen
  \bibfield  {author} {\bibinfo {author} {\bibfnamefont {M.}~\bibnamefont
  {Bellini}}, \bibinfo {author} {\bibfnamefont {C.}~\bibnamefont {Lyng\aa{}}},
  \bibinfo {author} {\bibfnamefont {A.}~\bibnamefont {Tozzi}}, \bibinfo
  {author} {\bibfnamefont {M.~B.}\ \bibnamefont {Gaarde}}, \bibinfo {author}
  {\bibfnamefont {T.~W.}\ \bibnamefont {H\"ansch}}, \bibinfo {author}
  {\bibfnamefont {A.}~\bibnamefont {L'Huillier}},\ and\ \bibinfo {author}
  {\bibfnamefont {C.-G.}\ \bibnamefont {Wahlstr\"om}},\ }\bibfield  {title}
  {\bibinfo {title} {Temporal coherence of ultrashort high-order harmonic
  pulses},\ }\href {https://doi.org/10.1103/PhysRevLett.81.297} {\bibfield
  {journal} {\bibinfo  {journal} {Phys. Rev. Lett.}\ }\textbf {\bibinfo
  {volume} {81}},\ \bibinfo {pages} {297} (\bibinfo {year} {1998})}\BibitemShut
  {NoStop}%
\bibitem [{\citenamefont {Sundaram}\ and\ \citenamefont
  {Niu}(1999)}]{PhysRevB.59.14915}%
  \BibitemOpen
  \bibfield  {author} {\bibinfo {author} {\bibfnamefont {G.}~\bibnamefont
  {Sundaram}}\ and\ \bibinfo {author} {\bibfnamefont {Q.}~\bibnamefont {Niu}},\
  }\bibfield  {title} {\bibinfo {title} {Wave-packet dynamics in slowly
  perturbed crystals: Gradient corrections and {{B}}erry-phase effects},\
  }\href {https://doi.org/10.1103/PhysRevB.59.14915} {\bibfield  {journal}
  {\bibinfo  {journal} {Phys. Rev. B}\ }\textbf {\bibinfo {volume} {59}},\
  \bibinfo {pages} {14915} (\bibinfo {year} {1999})}\BibitemShut {NoStop}%
\bibitem [{\citenamefont {Kilen}\ \emph {et~al.}(2020)\citenamefont {Kilen},
  \citenamefont {Kolesik}, \citenamefont {Hader}, \citenamefont {Moloney},
  \citenamefont {Huttner}, \citenamefont {Hagen},\ and\ \citenamefont
  {Koch}}]{PhysRevLett.125.083901}%
  \BibitemOpen
  \bibfield  {author} {\bibinfo {author} {\bibfnamefont {I.}~\bibnamefont
  {Kilen}}, \bibinfo {author} {\bibfnamefont {M.}~\bibnamefont {Kolesik}},
  \bibinfo {author} {\bibfnamefont {J.}~\bibnamefont {Hader}}, \bibinfo
  {author} {\bibfnamefont {J.~V.}\ \bibnamefont {Moloney}}, \bibinfo {author}
  {\bibfnamefont {U.}~\bibnamefont {Huttner}}, \bibinfo {author} {\bibfnamefont
  {M.~K.}\ \bibnamefont {Hagen}},\ and\ \bibinfo {author} {\bibfnamefont
  {S.~W.}\ \bibnamefont {Koch}},\ }\bibfield  {title} {\bibinfo {title}
  {Propagation induced dephasing in semiconductor high-harmonic generation},\
  }\href {https://doi.org/10.1103/PhysRevLett.125.083901} {\bibfield  {journal}
  {\bibinfo  {journal} {Phys. Rev. Lett.}\ }\textbf {\bibinfo {volume} {125}},\
  \bibinfo {pages} {083901} (\bibinfo {year} {2020})}\BibitemShut {NoStop}%
\bibitem [{\citenamefont {Jackson}(1998)}]{jackson_classical_1999}%
  \BibitemOpen
  \bibfield  {author} {\bibinfo {author} {\bibfnamefont {J.~D.}\ \bibnamefont
  {Jackson}},\ }\href@noop {} {\emph {\bibinfo {title} {Classical
  electrodynamics}}},\ \bibinfo {edition} {3rd}\ ed.\ (\bibinfo  {publisher}
  {Wiley},\ \bibinfo {address} {New York, {NY}},\ \bibinfo {year}
  {1998})\BibitemShut {NoStop}%
\bibitem [{\citenamefont {Jensen}\ and\ \citenamefont
  {Madsen}(2021)}]{PhysRevB.104.054309}%
  \BibitemOpen
  \bibfield  {author} {\bibinfo {author} {\bibfnamefont {S.~V.~B.}\
  \bibnamefont {Jensen}}\ and\ \bibinfo {author} {\bibfnamefont {L.~B.}\
  \bibnamefont {Madsen}},\ }\bibfield  {title} {\bibinfo {title} {Edge-state
  and bulklike laser-induced correlation effects in high-harmonic generation
  from a linear chain},\ }\href {https://doi.org/10.1103/PhysRevB.104.054309}
  {\bibfield  {journal} {\bibinfo  {journal} {Phys. Rev. B}\ }\textbf {\bibinfo
  {volume} {104}},\ \bibinfo {pages} {054309} (\bibinfo {year}
  {2021})}\BibitemShut {NoStop}%
\bibitem [{\citenamefont {Jensen}\ and\ \citenamefont
  {Madsen}(2020)}]{Jensen2020}%
  \BibitemOpen
  \bibfield  {author} {\bibinfo {author} {\bibfnamefont {S.~V.~B.}\
  \bibnamefont {Jensen}}\ and\ \bibinfo {author} {\bibfnamefont {L.~B.}\
  \bibnamefont {Madsen}},\ }\bibfield  {title} {\bibinfo {title} {Nondipole
  effects in laser-assisted electron scattering},\ }\href
  {https://doi.org/10.1088/1361-6455/aba718} {\bibfield  {journal} {\bibinfo
  {journal} {J. Phys. B}\ }\textbf {\bibinfo {volume} {53}},\ \bibinfo {pages}
  {195602} (\bibinfo {year} {2020})}\BibitemShut {NoStop}%
\bibitem [{\citenamefont {Lund}\ and\ \citenamefont {Madsen}(2021)}]{Lund2021}%
  \BibitemOpen
  \bibfield  {author} {\bibinfo {author} {\bibfnamefont {M.~M.}\ \bibnamefont
  {Lund}}\ and\ \bibinfo {author} {\bibfnamefont {L.~B.}\ \bibnamefont
  {Madsen}},\ }\bibfield  {title} {\bibinfo {title} {Nondipole photoelectron
  momentum shifts in strong-field ionization with mid-infrared laser pulses of
  long duration},\ }\href {https://doi.org/10.1088/1361-6455/ac20e2} {\bibfield
   {journal} {\bibinfo  {journal} {J. Phys. B}\ }\textbf {\bibinfo {volume}
  {54}},\ \bibinfo {pages} {165602} (\bibinfo {year} {2021})}\BibitemShut
  {NoStop}%
\bibitem [{\citenamefont {M\"ucke}(2011)}]{PhysRevB.84.081202}%
  \BibitemOpen
  \bibfield  {author} {\bibinfo {author} {\bibfnamefont {O.~D.}\ \bibnamefont
  {M\"ucke}},\ }\bibfield  {title} {\bibinfo {title} {Isolated high-order
  harmonics pulse from two-color-driven {{B}}loch oscillations in bulk
  semiconductors},\ }\href {https://doi.org/10.1103/PhysRevB.84.081202}
  {\bibfield  {journal} {\bibinfo  {journal} {Phys. Rev. B}\ }\textbf {\bibinfo
  {volume} {84}},\ \bibinfo {pages} {081202} (\bibinfo {year}
  {2011})}\BibitemShut {NoStop}%
\bibitem [{\citenamefont {Golde}\ \emph {et~al.}(2011)\citenamefont {Golde},
  \citenamefont {Kira}, \citenamefont {Meier},\ and\ \citenamefont
  {Koch}}]{Golde2011}%
  \BibitemOpen
  \bibfield  {author} {\bibinfo {author} {\bibfnamefont {D.}~\bibnamefont
  {Golde}}, \bibinfo {author} {\bibfnamefont {M.}~\bibnamefont {Kira}},
  \bibinfo {author} {\bibfnamefont {T.}~\bibnamefont {Meier}},\ and\ \bibinfo
  {author} {\bibfnamefont {S.~W.}\ \bibnamefont {Koch}},\ }\bibfield  {title}
  {\bibinfo {title} {Microscopic theory of the extremely nonlinear terahertz
  response of semiconductors},\ }\href
  {https://doi.org/https://doi.org/10.1002/pssb.201000840} {\bibfield
  {journal} {\bibinfo  {journal} {Phys. Status Solidi B}\ }\textbf {\bibinfo
  {volume} {248}},\ \bibinfo {pages} {863} (\bibinfo {year}
  {2011})}\BibitemShut {NoStop}%
\end{thebibliography}
\end{document}


\title{Supplemental Material \\ Propagation time and nondipole contributions to intraband high-harmonic generation}

\author{Simon Vendelbo Bylling Jensen}
\affiliation{Department of Physics and Astronomy, Aarhus
University, DK-8000 Aarhus C, Denmark}

\author{Lars Bojer Madsen}
\affiliation{Department of Physics and Astronomy, Aarhus
University, DK-8000 Aarhus C, Denmark}

\date{\today}

\renewcommand{\theequation}{S.\arabic{equation}}
\renewcommand{\thefigure}{S.\arabic{figure}}

\maketitle
\section{Discussion and Comparison to HHG in gases.}

In this section, we will briefly outline the interpretation of the emission process for HHG in gases and compare it to the intraband HHG process. For HHG in gases, the three-step model prescribes that only the contribution of the electron wavepacket which returns to the nucleus will emit harmonics at the recombination step \cite{PhysRevLett.68.3535,PhysRevLett.71.1994}. This happens as a part of the electron wavepacket returns to spatially overlap with the nucleus to form a time-dependent dipole moment and generate stimulated emission. When describing the observed spectrum at a detector, the signal will not only consist of a single atom radiation pattern but will consist of the coherent sum of the radiation from many atoms in the gas. The observed HHG signal is thus restricted on constructive interference, which infers that HHG is emitted mostly in the propagation direction of the driving field. For an account of such macroscopic effects of gaseous HHG see, e.g., Ref. \cite{Gaarde2008}. As a result of macroscopic effects, additional information regarding the generation process can be gained by considering the temporal and spatial distribution of the emitted radiation. For HHG in gases, this can be applied to, e.g., differentiate between long and short trajectories of the generation process \cite{PhysRevLett.81.297}. 

For the intraband generation process, the dynamics are widely different. Here the generation process is described from the point of view of Bloch states, which are spread in real space across many atomic sites. In the semiclassical model, a wavepacket of Bloch electrons is propagated. This wavepacket is spread across many lattice sites, however, localized when compared to the applied long-wavelength driving field. The current induced by such a wavepacket thus accounts for the coherent motion of electron contributions across many lattice sites. Furthermore, the intraband process does not require the recombination step making it intrinsically different from the gaseous case. In the intraband dynamics, the electron wavepacket induces a current throughout its trajectory, which causes the emission of harmonics. We expect, based on this widely different nature of the generation process, that macroscopic effects of another nature arise for HHG in solids. Of such effects are the propagational time delay, which is introduced as the electron wavepacket propagates during the intraband emission process. Propagation time delay is expected to be relevant for material characteristics of thin samples where other propagation effects are reduced. We expect, similar to the gaseous case, that such propagation effects imprint an additional layer of information regarding the generation process onto the observed spectra.   
 
\section{Analytical derivation based on the nondipole strong-field approximation}
\begin{figure}
\includegraphics[width=17.2 cm]{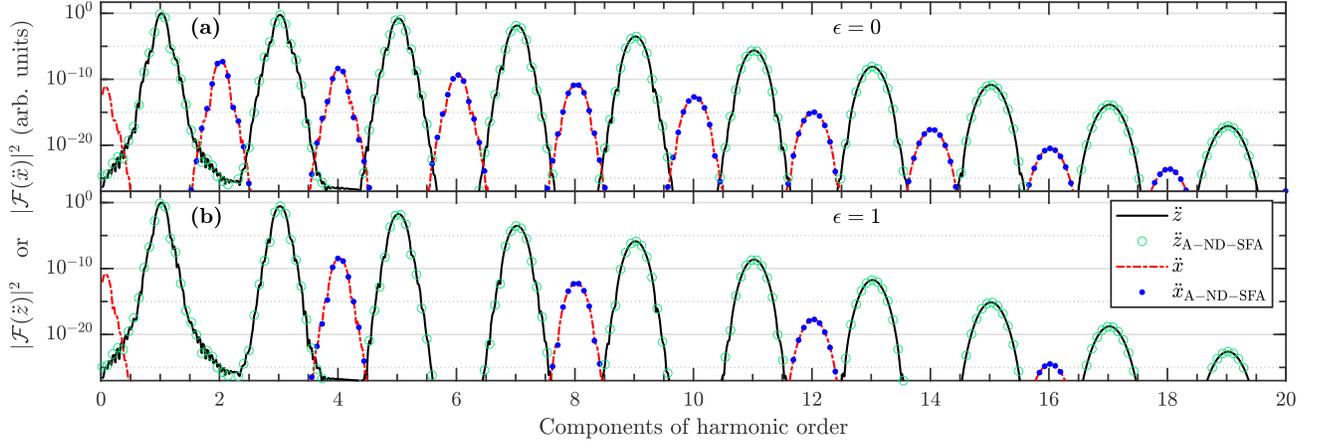}
\caption{Norm square of the Fourier transform of the acceleration along the polarization $\ddot{z}$ or propagation $\ddot{x}$ direction for a driving field with (a) linear $\epsilon = 0$ and (b) circular $\epsilon = 1$ polarization. The acceleration induced by the electromagnetic field and the $\text{A-ND-SFA}$ approach of Eq.~\eqref{eq:analytical} is compared with similar parameters as in Fig. 3 of the main text.} \label{fig:SM1}
\end{figure}

Here we derive the leading-order nondipole corrections to the acceleration of the electron wavepacket. The relevant quantities, i.e., parameters for the vector potential and sample are given in the main text of the paper, but to make the derivation easier to follow we have made the following presentation self-contained and will reintroduce most quantities.
We examine a spatial- and time-inversion symmetric sample, with three dimensional dispersion on the form
\begin{equation}
\varepsilon(\textbf{k}) = \frac{\hbar^2}{4 a^2 m_e} \left\lbrace 1 + \sum_n \sum_{i=\lbrace x,y,z\rbrace} c_{n,i} \cos(n k_i a) \right\rbrace, \label{bandstructure}
\end{equation}
with Fourier coefficients $c_{n,i}$ and lattice spacing $a$. An intraband electron wavepacket with center position $\bm r$ and wavevector $\bm k$ is described with the semiclassical equations \cite{PhysRevB.59.14915}
\begin{equation}
\hbar \dot{\bm k} = -e\left( \bm E + \dot{\bm r} \times \bm{B}\right) \ \ \ \  \ \ \ \ \mathrm{and}  \ \ \ \ \ \ \ \  \dot{\bm r} = \frac{1}{\hbar}\pdv{\varepsilon(\bm k)}{\bm k}. \label{rdot}
\end{equation}
The space- and time-dependent electric and magnetic fields $\bm E$ and $\bm B$ are derived from the vector potential $\bm{A} = A_0 ( \eta ) \left[0, \epsilon \cos(\eta), \sin(\eta) \right]^T$. Here $A_0 ( \eta )$ is an envelope function, and $\eta = \omega t - \omega x/c$ with the speed of light $c$. The polarization is defined from $\epsilon \in [0;1]$, with $\epsilon = 0$ and $\epsilon = 1$ being linearly and circularly polarized light, respectively. When considering leading-order nondipole effects of the light-matter interaction in the long-wavelength regime we employ notation similar to Ref.~\cite{PhysRevA.101.043408}. This is done by expanding the vector potential as $\bm A = \sum_{l=0}^\infty \bm A^{(l)}$ with $\bm A^{(l)}$ being the $l$'th order of $\omega x/c$, in which $\bm A^{(l)} = (l!)^{-1}  \dv*[l]{\bm A}{\eta} \eval{}_{\eta = \omega t} \left(-\omega x/c\right)^l$. The electric and magnetic fields inherit the notation of the $\bm A^{(l)}$ from which they are derived. Thereby we have $\bm E^{(l)}=-\partial_t \bm A^{(l)}$ and $\bm B^{(l)}=\nabla \times \bm A^{(l)}$. In the nondipole strong-field approximation of Ref.~\cite{PhysRevA.101.043408}, the dynamics induced for the polarization plane is found within the dipole approximation. In addition to this, the leading-order nondipole correction, originating from the interaction of the magnetic field on the dipole-induced trajectory is included. The leading-order approximation to Eq.~\eqref{rdot} hence reads 
 \begin{equation}
\hbar \dot{\bm k} = -e \left[ \left(- \partial_t \bm A^{(0)} \right) + \left(\dot{r}_y B_z^{(1)}-\dot{r}_z B_y^{(1)} \right)\widehat{x}\right]  \ \ \ \  \ \ \ \ \mathrm{and}  \ \ \ \ \ \ \ \  \dot{\bm r} = \frac{1}{\hbar}\pdv{\varepsilon(\bm k)}{\bm k}. \label{ndsfa}
\end{equation}
Here the dynamics in the polarization plane are determined from $\bm A^{(0)}$, the term within the dipole approximation, and in addition, a nondipole correction is included as the second term, which affects the dynamics in the propagation direction of the driving field. To solve Eq.~\eqref{ndsfa} analytically, we start by only considering the dynamics in the polarization plane. From the induced motion in the polarization plane, we can later determine the nondipole induced motion in the propagation direction. The motion in the propagation direction does not have any back-action on the motion in the polarization plane, due to $\bm A^{(0)}$ containing no spatial dependence. Thereby the problem can be separated into two steps, where we start by addressing only the motion in the polarization $(y,z)$-plane. To do this, we consider the dipole-approximated semiclassical equations where Eq.~\eqref{rdot} reduces to
\begin{equation}
\hbar \dot{\bm k} = e \partial_t \bm A^{(0)} \ \ \ \  \ \ \ \ \mathrm{and}  \ \ \ \ \ \ \ \  \dot{\bm r} = \frac{1}{\hbar}\pdv{\varepsilon(\bm k)}{\bm k}. \label{rdotdip}
\end{equation}
The first equation is readily solved analytically for $\bm k$, as $\bm k = \frac{e}{\hbar} \bm A^{(0)} + \bm k_0$. We select the wavepacket to originate at the $\Gamma$ point, $\bm k_0 = [0,0,0]^T$. When solving the last part of Eq.~\eqref{rdotdip}, the obtained $\bm k$ is inserted with the bandstructure defined from Eq.~\eqref{bandstructure} to give
\begin{equation}
\dot{\bm r}  = -\frac{\hbar}{4 a m_e}  \sum_n \left( \begin{bmatrix} 0 \\ n c_{n,y}\sin(\frac{n a e}{\hbar} \epsilon A_0(\omega t)  \cos(\omega t) ) \\ n c_{n,z} \sin(\frac{n a e}{\hbar} A_0(\omega t) \sin(\omega t) )\end{bmatrix} \right).
\end{equation}
We identify the following relations, 
\begin{align}
\sin(z \cos(\theta)) &= 2 \sum_{k=0}^\infty (-1)^k  J_{2k+1} (z) \cos((2k +1)\theta), \label{sincos}\\
\sin(z \sin(\theta)) &= 2 \sum_{k=0}^\infty  J_{2k+1} (z) \sin((2k +1)\theta), \label{sinsin}
\end{align}
to find
\begin{equation}
\dot{\bm r}  =  -\frac{\hbar}{2 a m_e}   \sum_n \sum_{l=1,3,5,...}^\infty  \begin{bmatrix} 0 \\  n c_{n,y} (-1)^{(l-1)/2}  J_{l} (\frac{n a e}{\hbar} A_0(\omega t) \epsilon) \cos(l \omega t) \\ n c_{n,z} J_{l} (\frac{n a e}{\hbar} A_0(\omega t)) \sin(l\omega t) \end{bmatrix}.  \label{eq:diprdot}
\end{equation}
Returning to Eq.~\eqref{ndsfa}, the motion in the polarization plane has now been obtained. We now solve Eq.~\eqref{ndsfa} in the propagation direction starting with
\begin{equation}
\hbar \dot{k_x} = -e \left[\dot{r}_y B_z^{(1)}-\dot{r}_z B_y^{(1)}\right] \widehat{x}. \label{kdot2}
\end{equation}
To solve this, we start by giving the dominant contribution to the magnetic field, where we apply the slowly varying envelope approximation $\pdv{A_0( \eta )}{x} \approx 0$, such that
\begin{equation}
\bm B^{(1)} = \curl{\textbf{A}^{(1)}} = \begin{bmatrix} 
		0  \\ 
		A_0( \omega t ) \frac{\omega}{c} \cos(\omega t) + \partial_{\eta} A_0( \eta )\mid_{\eta = \omega t} \frac{\omega}{c} \sin(\omega t)  \\ 
		  A_0( \omega t ) \epsilon \frac{\omega}{c} \sin(\omega t)  - \partial_{\eta} A_0( \eta )\mid_{\eta = \omega t} \epsilon \frac{\omega}{c} \cos(\omega t)
	\end{bmatrix} \approx  \begin{bmatrix} 
		0  \\ 
		A_0(\omega t) \frac{\omega}{c} \cos(\omega t)  \\ 
		  A_0(\omega t) \epsilon \frac{\omega}{c} \sin(\omega t) 
	\end{bmatrix} \label{magnetic_field}
\end{equation}
Inserting this along with Eq.~\eqref{eq:diprdot}, and applying trigonometric identities as well as the recurrence relations of the Bessel functions 
\begin{equation}
\frac{2k}{x} J_k (x) = J_{k+1}(x) + J_{k-1} (x),
\end{equation}
one obtains that Eq.~\eqref{kdot2} reduces to
\begin{align}
\dot{k_x}  = -\frac{\hbar \omega}{2 a^2 m_e c}  \sum_n  \sum_{l=2,4,6...}^\infty  \left\lbrace l \sin(l \omega t) \left[  c_{n,z} J_{l} \left(\frac{n a e}{\hbar} A_0(\omega t)\right) + c_{n,y} (-1)^{l/2} J_{l} \left(\frac{n a e \epsilon}{\hbar} A_0(\omega t)\right) \right]\right\rbrace. \label{k_xdot}
\end{align}
We find the differentiated crystal momenta in the propagation direction to be of order $1/c$ less than the differentiated crystal momenta in the polarization plane. Therefore, we expand the bandstructure in a constant effective mass approach for the propagation direction, which is proficient for $k_x \ll 1$. In such approximation the effective mass $m^\star$ in the propagation direction is defined from
\begin{equation}
\pdv{\varepsilon(\textbf{k})}{\textbf{k}} \approx \frac{\hbar^2 (\textbf{k}-\textbf{k}_0)}{m^\star},
\end{equation}
which for the dispersion of Eq.~\eqref{bandstructure} along the $x$-direction gives
\begin{equation}
\pdv{\varepsilon(\textbf{k})}{k_x} \approx -\frac{\hbar^2}{4m_e}  \sum_n \left(c_{n,x} n^2  \right) k_x,
\end{equation}
defining the effective mass along the $x$-direction as
\begin{equation}
m^\star = -\frac{4m_e}{\sum_n \left(c_{n,x} n^2  \right)}.
\end{equation}
We choose the notation of an effective mass instead of directly using the Fourier coefficients due to two considerations (i) To remind that an approximation is made for the dispersion in the propagation direction, whereas the dispersion in the polarization plane is not approximative. (ii) To remind that the dispersion along the propagation direction might be very different from that of the polarization plane, and thus the approach at hand can be generalized for other material structures. 
For the propagation direction with $k_x \ll 1$ the constant effective mass approach allows us to approximate the last part of Eq.~\eqref{ndsfa} with
\begin{equation}
\ddot{r}_x  = \frac{\hbar }{m^\star}  \dot{k}_x, 
\end{equation}
such that
\begin{equation}
\ddot{r}_x = -\frac{\hbar^2 \omega}{2 a^2 m_e m^\star c}  
\sum_n  \sum_{l=2,4,6...}^\infty  \left\lbrace l \sin(l \omega t) \left[  c_{n,x}  J_{l} \left(\frac{n a e}{\hbar} A_0(\omega t)\right) +  c_{n,y} (-1)^{l/2} J_{l} \left(\frac{n a e \epsilon}{\hbar} A_0(\omega t)\right) \right]\right\rbrace . \label{Eharm}
\end{equation}
Combining the equations of motion for the intraband wavepacket within the nondipole strong-field approximation, we find the wavepacket acceleration to be
\begin{equation}
\ddot{\bm r} = \frac{\hbar \omega}{2 a m_e}  \sum_n  \sum_{l=1,3,5,...}^\infty \begin{bmatrix}  -\frac{\hbar (l+1)}{a m^\star c}  \left\lbrace   c_{n,z}  J_{l+1} \left(\frac{n a e}{\hbar}A_0(\omega t) \right) +  c_{n,y} (-1)^{(l+1)/2} J_{l+1} \left(\frac{n a e}{\hbar}\epsilon A_0(\omega t) \right) \right\rbrace  \sin((l+1) \omega t) \\ nl  c_{n,y} (-1)^{(l-1)/2}  J_{l} (\frac{n a e }{\hbar} \epsilon A_0(\omega t)) \sin(l \omega t) \\ -nl  c_{n,z} J_{l} (\frac{n a e}{\hbar} A_0(\omega t)) \cos(l\omega t) \end{bmatrix}. \label{eq:analytical}
\end{equation}
This analytical result, based on the nondipole strong-field approximation, will be denoted with A-ND-SFA. It accurately describes the dynamics of the intraband electron wavepacket, as observed in Fig.~\ref{fig:SM1} as it captures all relevant nondipole corrections. It fails to capture the zeroth harmonic, which arises due to the variations of the envelope function and was neglected in Eq.~\eqref{magnetic_field}. The analytical assessment of nondipole corrections to the acceleration, made possible by the leading-order approximation of Eq.~\eqref{ndsfa}, allows us to further proceed to include propagation time delay effects and describe the observed spectra. This is done in the main text in relation to Fig. 4. 

We can further proceed to derive selection rules for the harmonic components of the acceleration. First a slowly varying envelope approximation is made $A_0(\omega t) = A_0$, in which the Bessel functions $J_l$ will remain constant through time. Secondly, we shorten the notation by introducing the Bloch frequency, $\omega_B = eaA_0\omega/\hbar$ to find
\begin{equation}
\ddot{\bm r} = \frac{\hbar \omega}{2 a m_e}  \sum_n  \sum_{l=1,3,5,...}^\infty  \begin{bmatrix}  -\frac{\hbar (l+1)}{a m^\star c}  \left\lbrace   c_{n,z}  J_{l+1} \left(\frac{n \omega_B}{\omega} \right) +  c_{n,y} (-1)^{(l+1)/2} J_{l+1} \left(\frac{n \omega_B \epsilon }{\omega} \right) \right\rbrace  \sin((l+1) \omega t) \\ nl  c_{n,y} (-1)^{(l-1)/2}  J_{l} (\frac{n \omega_B \epsilon }{\omega}) \sin(l \omega t) \\ -nl  c_{n,z} J_{l} (\frac{n \omega_B}{\omega}) \cos(l\omega t) \end{bmatrix}. \label{eq:analytical2}
\end{equation}
To reveal the selection rules, we perform a Fourier transform with Fourier variable $\Omega > 0$ and find
\begin{align}
\ddot{\bm r}(\Omega) &=  \frac{1}{\sqrt{2\pi}} \int_{-\infty}^\infty dt \  e^{-i\omega t } \ddot{\bm r}(t)\\
\ddot{\bm r}(\Omega) &=  \frac{\hbar \omega \sqrt{2\pi}}{4 a m_e}  \sum_n  \sum_{l=1,3,5,...}^\infty  \begin{bmatrix}  -i\frac{\hbar (l+1)}{a m^\star c}  \left\lbrace   c_{n,z}  J_{l+1} \left(\frac{n \omega_B}{\omega} \right) +  c_{n,y} (-1)^{(l+1)/2} J_{l+1} \left(\frac{n \omega_B \epsilon }{\omega} \right) \right\rbrace  \delta\left(\Omega-(l+1)\omega\right)  \\ inl  c_{n,y} (-1)^{(l-1)/2}  J_{l} (\frac{n \omega_B \epsilon }{\omega})   \delta\left(\Omega-l\omega\right) \\ -nl  c_{n,z}  J_{l} (\frac{n \omega_B}{\omega})  \delta\left(\Omega-l\omega\right) \end{bmatrix}. \label{eq:analytical3}
\end{align}
From Eq.~\eqref{eq:analytical3} the selection rules can be established by considering the Dirac delta functions in each dimension. For the $y$- and $z$-direction we have $\delta(\Omega - l\omega )$. This will for $l=1,3,5, ...$ only allow for odd frequency components as $\Omega = l \omega = 1 \omega, 3\omega, 5\omega ...$. For the nondipole harmonics along the $x$-diction the Dirac delta function is $\delta(\Omega -(l+1)\omega)$. This will for  $l=1,3,5,...$ only allow for frequencies of even orders since $\Omega = (l+1)\omega = 2\omega, 4\omega, 6\omega, ...$. Regarding these nondipole selection rules, it is worth noting that for the $c_{n,z} = c_{n,y}$ and $\epsilon=1$ the Bessel functions in the $x$-direction of Eq.~\eqref{eq:analytical3} reduces 
\begin{equation}
  J_{l} \left(\frac{n \omega_B}{\omega} \right) +  (-1)^{l/2} J_{l} \left(\frac{n \omega_B }{\omega} \right) =
    \begin{cases}
       2 J_{l} \left(\frac{n \omega_B}{\omega} \right) & \text{for } l = 4,8,12, ...\\
      0 & \text{for } l = 2,6,10, ...
    \end{cases}.       
\end{equation}
This causes a fourth-order selection rule for the nondipole acceleration if the dispersion in both directions in the polarization plane are identical $c_{n,z} = c_{n,y}$ and a circular polarized electric field $\epsilon=1$ is applied. If one wishes, the spectra in the long-pulse limit can be obtained from Eq.~\eqref{eq:analytical3} applying Eq. (3) in the main text. The intensity of each harmonic component can thus be estimated through the square modulus of the Fourier transform. From the modulated Dirac comb of Eq.~\eqref{eq:analytical3} the peak height for the oscillation amplitude can be approximated in each direction as
\begin{align}
I_{l,x} &\propto \delta_{l,\mathrm{even}}    \left[\frac{\omega l}{a^2 m^\star c}\right]^2 \left|\sum_n \left[ c_{n,z}  J_{l} \left(\frac{n \omega_B}{\omega} \right) +  c_{n,y} (-1)^{l/2} J_{l} \left(\frac{n \omega_B \epsilon }{\omega} \right) \right]  \right|^2, \\
I_{l,y} &\propto \delta_{l,\mathrm{odd}}    \left[\frac{\omega l}{a}\right]^2 \left|\sum_n \left[n   c_{n,y} J_{l} \left(\frac{n \omega_B \epsilon }{\omega} \right) \right] \right|^2, \\
I_{l,z} &\propto \delta_{l,\mathrm{odd}}    \left[\frac{\omega l}{a}\right]^2 \left|\sum_n \left[n   c_{n,z} J_{l} \left(\frac{n \omega_B }{\omega} \right)\right]  \right|^2, 
\end{align}
Here we immediately see that the nondipole contribution is of order $(a m^\star c)^{-2}$ lower than the dipole induced acceleration. We thus expect nondipole induced harmonics to be of increasing importance for decreasing effective mass along the propagation direction. It is furthermore clear that the nondipole dynamics can be directly applied to gain access to the symmetry of the dispersion in the polarization plane since the nondipole dynamics depend on the interference between the Bessel functions describing the motion along each direction of the polarization plane.